\documentclass[aps,prd,superscriptaddress,groupedaddress,nofootinbib,preprint,longbibliography
]{revtex4-1}
\usepackage{asymptote} 
\usepackage{amsmath,amssymb,amsfonts}
\usepackage{color}
\usepackage{xcolor}
\definecolor{red}{rgb}{1,0,0}
\usepackage[colorlinks=true, linkcolor=blue, urlcolor=magenta, citecolor=red]{hyperref}
\usepackage{graphicx}
\graphicspath{{./pic/}}
\newcommand{\bw}{\begin{widetext}}
\newcommand{\ew}{\end{widetext}}
\newcommand{\be}{\begin{equation}}
\newcommand{\en}{\end{equation}}
\newcommand{\bee}{\begin{equation}}
\newcommand{\ene}{\end{equation}}
\newcommand{\bea}{\begin{eqnarray}}
\newcommand{\ena}{\end{eqnarray}}
\newcommand{\bes}{\begin{subequations}}
\newcommand{\ens}{\end{subequations}}
\newcommand{\bef}{\begin{figure}}
\newcommand{\enf}{\end{figure}}

\def\thefootnote{\fnsymbol{footnote}}

\def\ie{{\it i.e.}}

\def\cala{\mathcal{A}}

\def\call{\mathcal{L}}

\def\calo{\mathcal{O}}

\def\calt{\mathcal{T}}

\begin{document}
%\linenumbers

%%%%%%%%%%%%%%%%%%%%%%%%%%%%%%%%%%%%%%%%%%%%%
%\title{Electromagnetic Dynamics of Spinor Particle in a Space with Spin Noncommutativity of Coordinates}
%\title{Spin Dynamics and Spin-Hall Conductivity on a Coordinate Dependent Noncommutative space}
\title{Chiral Spin Noncommutative Space and Anomalous Dipole Moments}
\author{Kai Ma}
\email[Electronic address: ]{makainca@yeah.net}
\affiliation{School of Physics Science, Shaanxi University of Technology, Hanzhong 723000, Shaanxi, People's Republic of China}

\date{\today}

\begin{abstract}
We introduce a new model of spin noncommutative space in which noncommutative extension of the coordinate operators are assumed to be chirality dependent. Noncommutative correspondences of classical fields are defined via Weyl ordering, and the maps are represented by a spin-dependent translation operator. Based on the maps, gauge field theory in chiral spin noncommutative space is established. The corresponding gauge transformations are induced by a local phase rotation on commutative functions, and hence are consistent with the ordinary gauge transformations by definition. Furthermore, a general extension of the ordering between Dirac matrix and gauge potential is introduced. Noncommutative corrections on equations of motion of matter and gauge fields are studied. We find that there are two kinds of corrections. The first kind of correction involves derivatives of the matter field, and hence contribute the ordinary Noether current. On the other hand, the second kind of correction depends only on derivatives of the gauge fields, and hence contribute anomalous magnetic and electric dipole moments of the matter field. Moreover, experimental bounds on the noncommutative parameters for muon lepton are studied in a simplified model. 
\end{abstract}

\maketitle

\tableofcontents

%%%%%%%%%%%%%%%%%%%%%%%%%%%%%%%%%%%%%%%%%%%%%%%%%%%%%%%%%%%%%%%%%%%%%%%%%%%%%%%%%%%%%%%%%%%%%%%%%%%%%%%%%%%%%%%%%%%%%%%%%
%
\setcounter{page}{1}
\renewcommand{\thefootnote}{\arabic{footnote}}
\setcounter{footnote}{0}

\section{Introduction}\label{sec:intro}
It is well-known that the geometry of the background spacetime is dynamical, and the General Relativity  (GR) is a concrete model for describing the gravity. However, it is also known that GR is inconsistent with the Quantum Mechanics (QM). Many models have been proposed to combine the GR and QM in a consistent way. And a noncommutative geometry appears as an effective phenomenological model of the background spacetime below the Planck scale~\cite{Freidel:2005me, Moffat:2000gr, Moffat:2000fv, Faizal:2013ioa}. 
A spacetime with noncommutative geometry is characterized by deformation of the fundamental algebra of position operators in ordinary QM. A large variety of noncommutative models have been proposed. For instance, the so called ``canonical noncommutativity" which is parameterized by a totally anti-symmetric constant tensor $\theta_{\mu\nu}$ as follows,
\bee\label{eq:def:ncx:c}
\big[ \tilde{x}_{\mu}\,, \tilde{x}_{\nu}] = i \theta_{\mu\nu}\,,
\ene
can appear when a D-brane is in a constant Neven-Schwarz $B$-field~\cite{Seiberg:1999vs}. Such a noncomnutative algebra was also found in real quantum system~\cite{Gamboa:2000yq, Ho:2001aa}.

However, due to the constant property of the parameter $\theta_{\mu\nu}$, two preferential directions have to be given in advance. As consequences, rotational symmetry is broken~\cite{Douglas:2001ba, Szabo:2001kg, Ma:2017fnt}. Furthermore, the unitarity can also be violated if the temporal components are nonzero~\cite{Chaichian:2001pw,Chaichian:2000hy,Li:2005mi,Chaichian:2008gf}. Both dynamical and kinematical aspects of the above algebra have been extensively studied. For instance, distortions of energy levels of atoms~\cite{Chaichian:2000si,Zhang:2004yu,Zhang:2011kr,Zhang:2011zua,Ma:2017rwg,Ramirez:2017pmp}, contributions to the topological phase effects~\cite{Chaichian:2000hy,Chaichian:2001pw,Ma:2016rhk,Ma:2016vac,Ma:2014tua,Rodriguez:2017iti,Fateme:2015eaa,Wang:2017gyy}, corrections on the spin-orbital interactions~\cite{Ma:2011gc,Deriglazov:2016mhk,Wang:2018zaw,Deriglazov:2015wde,Deriglazov:2018vwa,Ren:2018rku}, as well as deformations of quantum speeds of relativistic charged particles~\cite{Wang:2017azq,Wang:2017arq,Deriglazov:2015zta,Guzman-Ramirez:2019ijo}.
Factually, long before studies on the canonical noncommutative model, Snyder has developed a Lorentz invariant noncommutative model which is defined by following algebra~\cite{Snyder:1946qz,Snyder:1947nq},
\bee\label{eq:def:ncx:Synder}
\big[ \tilde{x}_{\mu}\,, \tilde{x}_{\nu}] = i l^{2} L_{\mu\nu}\,,
\ene
where $L_{\mu\nu}$ are generators of the Lorentz group, and $l$ is a parameter with dimension of length and describes the magnitude of spacetime noncommutativity. It was further pointed out that the translation symmetry have to be broken~\cite{Yang:1947ud}. On the other hand, more and more models have been proposed for restoring the rotational invariance. For instance, in Ref.~\cite{Doplicher:1994tu}, the spacial components $\theta_{ij}$ are assumed to be operators commuting with each other and transforming as components of tensor.

In consideration of that spin degrees of freedom are representations of the Lorentz group, the right-hand side of \eqref{eq:def:ncx:Synder} was extended as the spin operator in Ref.~\cite{Falomir:2009cq}, and known as ``spin noncommutativity". In such a model, the spacetime and spin are coupled to each other, and hence a supersymmetric extensions of harmonic oscillator exist~\cite{Falomir:2009cq}. It was also shown that there is a strong anisotropy at small distance in Aharonov-Bohm scattering, even through the topological properties does not change~\cite{Das:2011tj}. 
%Furthermore, logarithmic dependence of $s$-levels energy for hydrogen atom was obtained in Ref.~\cite{xxx}. 
On the other hand, the spin noncommutative model introduced in Ref.~\cite{Falomir:2009cq} was extended to relativistic situation as follows,
\bea
\label{eq:def:ncx:spin}
\big[ \tilde{x}^{\mu} \,, \tilde{x}^{\nu}  \big]
&=& 
2i\theta\epsilon^{\mu\nu\alpha\beta}\sigma_{\alpha\beta} 
- i\theta^{2}\epsilon^{\mu\nu\alpha\beta}W_{\alpha}p_{\beta}\,,
\\[2mm]
\big[ \tilde{x}^{\mu} \,, \gamma^{\nu}  \big] 
&=&
-\frac{i}{2}\theta\epsilon^{\mu\nu\alpha\beta}p_{\alpha}\gamma_{\beta}\,,
\ena
where $W^{\mu} = \epsilon^{\mu\nu\alpha\beta} \sigma_{\nu\alpha}p_{\beta}$ is the Pauli-Lubansky pseudo-vector. Apparently, the above definitions are covariant, and hence the Lorentz symmetry was preserved. Such a phenomanological model can be found when we consider a spinning particle in curved spacetime~\cite{Ramirez:2013xga,Deriglazov:2015bqa}. Deformation of the ordinary Dirac equation has also been studied~\cite{Ferrari:2012bv}. However, it was pointed out that the micro-causality is violated~\cite{Huang:2012kf}. Furthermore, one can see that the above extension is nonlocal. In stead of utilizing the Pauli-Lubansky pseudo-vector which involves momenta operator, by assuming that the right-hand side of \eqref{eq:def:ncx:spin} is proportional to Dirac matrix directly, a new model was introduced in Ref.~\cite{Vasyuta:2016gyz}.

In this paper, we consider a more general class of spin noncommutative model, in which the noncommutative parameters are tensorial and more importantly are chiral-dependent. The contents of this paper is organized as follows: in Sec.~\ref{sec:tensorialSpinNC} we introduce our model by defining the maps from commutative fields (operators) to noncommutative fields (operators); in Sec.~\ref{sec:LocalGaugeTheory} we will study the local gauge theory, and give the gauge transformation. The equation of motion of the matter and gauge fields will be derived, and the noncommutative corrections on the Noether current and electric (magnetic) dipole moments will be investigated; summary and our conclusions will be given in the final section, Sec.~\ref{sec:sum}.

%%%%%%%%%%%%%%%%%%%%%%%%%%%%%%%%%%%
\section{Chiral Spin Noncommutativity}\label{sec:tensorialSpinNC}
Even through the commutators between coordinate operators defined in Refs.~\cite{Falomir:2009cq,Deriglazov:2015bqa,Ferrari:2012bv} are different, the spin noncommutative spacetime can be introduced uniformly by a map from ordinary coordinates $x^{\mu}$ to the noncommutative coordinates $\tilde{x}^{\mu}$ as follows,
%The general spin noncommutativity can be described by following map from ordinary position operator $x^{\mu}$ to noncommutative one $X^{\mu}$,
\bee\label{eq:generalSpinNC}
\tilde{x}^{\mu} = x^{\mu} + \Gamma_{\mu}\left( \{\theta\}; x, p \right)\,,
\ene
where $\Gamma_{\nu}\left( \{\theta\}; x, p \right)$ is a vector function with nontrivial spinor structures (\ie, combination of $\gamma$-matrix) of the commutative coordinate and momenta operators, and parameterized by a set of noncommutative parameters $\{\theta\}$. It is also worthy to note that, the unit of the parameters are somehow model dependent. In this paper we study the case of that $\Gamma_{\mu}$ is proportional to the Dirac matrix $\gamma_{\mu}$, but parameterized by two tensorial noncommutative parameters for left- and right-handed chiral spinors, respectively.

%where $\theta^{\mu\nu}$ is a constant parameter encoding the strongth of the noncommutativity, and $S_{\nu}$ denmostrates the spinor structures. When the noncommutative parameter $\theta^{\mu\nu}$ is proportional to the metric $\eta^{\mu\nu}$, say, $\theta^{\mu\nu} = \theta\eta^{\mu\nu}$, the map given in \eqref{eq:generalSpinNC} reduce to maps in Ref.~\cite{Gomes:2010xk} and Ref.~\cite{Vasyuta:2016gyz} with $S_{\nu}=W_{\nu}$ and $S_{\nu}=i\gamma_{\nu}$, respectively. In this sense, the parameter $\theta^{\mu\nu}$ describes the geometric properties of the underlying space. By assuming that momenta in noncommutative space are the same as in the commutative case, it is easy to write down the full closed algebra. The noncommutative corresponding of any ordinary function $f(x)$ can be obtained by using the Weyl ordering

\subsection{Position Operator with Chirality}
The spin noncommutativity studied in this paper is defined by following map,
\bee\label{eq:generalSpinNC}
\tilde{x}^{\mu} = x^{\mu} + \theta^{\mu\nu}_{L}\gamma_{\nu}\gamma_{L} + \theta^{\mu\nu}_{R}\gamma_{\nu}\gamma_{R}\,,
\ene
where $\theta^{\mu\nu}_{L/R}$ are the noncommutative parameter. In our model, the noncommutative parameters $\theta^{\mu\nu}_{L/R}$ have dimensions of length, which is different from the canonical noncommutative model that is parameterized by a constant tensorial parameter having dimension of area. When the noncommutative parameter $\theta^{\mu\nu}_{L/R}$ is proportional to the metric $\eta^{\mu\nu}$, say, $\theta^{\mu\nu}_{L/R} = \theta_{L/R}\eta^{\mu\nu}$, the map given in \eqref{eq:generalSpinNC} is reduced to the one defined in Ref.~\cite{Vasyuta:2016gyz}. 
%In this sense, the parameter $\theta^{\mu\nu}$ describes the geometric properties of the underlying space. 
%By assuming that momenta in noncommutative space are the same as in the commutative case, it is easy to write down the full closed algebra. 

The above definition involves non-trivial matrix structures in spinor space, as a consequence, the noncommutative operator $\tilde{x}$ is not Hermitian in general. Even though that Hermitian is a necessary condition for an operator to be observable in standard quantum mechanics, it is not really necessary when the operator is defined in relativistic spinor space where the condition $\calo^\dag=\gamma_{0}\calo\gamma_{0}$ is more useful to obtain the conjugate Dirac equation and a real Lagrangian density~\cite{Greiner:1990tz}. We will use this condition in this paper, see Ref.~\cite{Bagchi:2009wb,Fring:2010pw} for more sophistic interpretations of this problem. One can easily find that $ \theta^{\mu\nu}_{L}$ and $\theta^{\mu\nu}_{R}$ have to be real for satisfying this condition.

In case of that the momenta operator in noncommutative space is assumed to be the same as the one in commutative space, \ie, $\tilde{p}_{\mu}=p_{\mu}=i\partial_{x^\mu}$, the full algebras are given as
\bea
\big[ \tilde{x}^{\mu} \,, \tilde{x}^{\nu}  \big]
&=& - i \big( \theta^{\mu\alpha}_{L}\theta^{\nu\beta}_{R} + \theta^{\mu\alpha}_{R}\theta^{\nu\beta}_{L}  \big) \sigma_{\alpha\beta} 
+ \gamma_{5} \big( \theta^{\mu\delta}_{L}\theta^{\nu}_{R\delta} - \theta^{\mu\delta}_{R}\theta^{\nu}_{L\delta}  \big) \,,
\\[2mm]
\big[ \tilde{p}^{\mu} \,, \tilde{p}^{\nu}  \big] &=& 0 \,,
\\[2mm]
\big[ \tilde{p}^{\mu} \,, \tilde{x}^{\nu}  \big] &=& i\eta^{\mu\nu}\,,
\\[2mm]
\big[ \tilde{x}^{\mu} \,, \sigma^{\alpha\beta}  \big] &=& 
-2i \big( \theta_{V}^{\mu\delta} + \theta_{A}^{\mu\delta}\gamma_{5} \big) 
\big( \gamma^{\alpha}\eta^{\beta}_{\;\;\delta} - \gamma^{\beta}\eta^{\alpha}_{\;\;\delta} \big) \,,
\\[2mm]
\big[ \tilde{p}^{\mu} \,, \sigma^{\alpha\beta}  \big] &=& 0\,,
\\[2mm]
\big[ \sigma^{\mu\nu} \,, \sigma^{\alpha\beta}  \big] 
&=& 
i\big( \eta^{\mu\alpha}\sigma^{\nu\beta} - \eta^{\nu\alpha}\sigma^{\mu\beta} + \eta^{\nu\beta}\sigma^{\mu\alpha} - \eta^{\mu\beta}\sigma^{\nu\alpha}  \big)\,.
\ena

\subsection{Weyl Ordering and Star-Product}
As the usual way, in order to have a dynamical theory on the noncommutative spacetime, definitions of the fields in terms of noncommutative coordinates, as well as actions of the noncommutative coordinates operators on functions defined on commutative spacetime have to be given unambiguously. This can be addressed by using the well-known methods Weyl ordering and Moyal-Groenewald product, respectively. For an ordinary function $f(x)$, its noncommutative correspondence, $f( \tilde{x})$,  is given via the Weyl ordering as 
\bee
f( \tilde{x} ) = \int \frac{ d^{4} k }{ (2\pi)^4 } F (k) e^{- i k_{\mu} \tilde{x}^{\mu}} \,,
\ene
where $ F(k)$ is the Fourier transformation of the function $f(x)$. It is worthy to point out that, because the function $f(x)$ can in general have non-trivial spinor structures, the ordering between the Fourier transformed function $F (k) $ and the exponential $e^{- i k_{\mu} \tilde{x}^{\mu}}$ in the integrand of above definition is also important. Furthermore, as the usual case, the noncommutive correspondence of an ordinary function $f(x)$ can be obtained by a translation operator as follows,
\bea
f( \widetilde{x} ) &=& \left[ \overrightarrow{\calt}( \theta_{L}, \theta_{R} )  f(x) \right] = \widetilde{f( x )} \,,
\\[3mm]
\overrightarrow{\calt}( \theta_{L}, \theta_{R} )  &=& 
 \exp{ \bigg[ 
\gamma_{\nu} \big(\theta^{\mu\nu}_{L}\gamma_{L} + \theta^{\mu\nu}_{R}\gamma_{R} \big) \overrightarrow{\partial}_{x^{\mu}}
\bigg] } \,.
\ena
Here for convenience, we have used the arrow ``$\overrightarrow{}$" to denote direction of the differential operator (rather than stand for a vector). As expected, the translation operator $\overrightarrow{ \calt}$ also satisfies the condition $\calo^\dag=\gamma_{0}\calo\gamma_{0}$, 
\ie,
$\overrightarrow{ \calt}^{\dag} =
\gamma^{0} \overleftarrow{\calt } \gamma^{0}
$.
This property is important for obtaining the noncommutative correspondence of the conjugate of a spinor function $\psi$, which can be written as
$
\overline{\widetilde{\psi}} 
%= \left( \overrightarrow{\calt} f(x) \right)^{\dag} \gamma_{0}
= \overline{ \psi } \,\overleftarrow{\calt }
$. 

As in the canonical noncommutative model, product of two noncommutative function can be represented by the Moyal-Groenewald star product. In our case, the star product can be defined as $\star = \overleftarrow{\calt } \; \overrightarrow{ \calt}$. For a spinor function $\psi(x)$ and its conjugate $\overline{\psi}(x)$, product of their noncommutative correspondences can be written as,
\bee
\overline{\widetilde{\psi( x )}}\, \widetilde{\psi( x )}  
= \overline{\psi}( \widetilde{x} ) \psi( \widetilde{x} ) 
= \overline{\psi}( x ) \star \psi( x ) \,.
\ene
Since the total probability have to be $1$, above product has to satisfy following condition,
\bee\label{eq:Star:TProb}
\int d^{4}x\; \overline{\psi}( x ) \star \psi( x ) = \int d^{4}x\; \overline{\psi}( x ) \psi( x )\,. 
\ene
This can be easily prove by using following equation,
\bee\label{eq:Star:TT}
\overline{\psi}( x ) \star \chi( x )
=
\overline{\psi}( x ) \overrightarrow{\calt}( -\theta_{L}, -\theta_{R}  ) \overrightarrow{\calt}( \theta_{L}, \theta_{R} ) \chi( x ) + \partial_{\mu}\Omega^{\mu}\,,
\ene
where $\overline{\psi}(x)$ and $\chi(x)$ are two arbitrary spinor functions, and $\partial_{\mu}\Omega^{\mu}$ is a total derivative that is not important in our case. It is apparently that the equation \eqref{eq:Star:TProb} holds in general. Furthermore, by using the relation \eqref{eq:Star:TT}, one can easily show that the condition \eqref{eq:Star:TProb} also holds under a local gauge transformation $\psi'(x)=e^{i\alpha(x)}\psi(x)$,
%\bee
%\int d^{4}x\; \overline{\psi'}( x ) \star \psi'( x ) 
%=
%\int d^{4}x\; \overline{\psi'}( x ) e^{-i\alpha(x)} \overleftarrow{T}( \theta_{L}, \theta_{R} )  \overrightarrow{T}( \theta_{L}, \theta_{R} ) e^{i\alpha(x)}\psi'( x ) 
%\ene
%By using the relation we have
\bee
\int d^{4}x\; \overline{\psi'} \star \psi'
=
\int d^{4}x\; \overline{\psi} e^{-i\alpha} \overrightarrow{T}( -\theta_{L}, -\theta_{R} )  \overrightarrow{T}( \theta_{L}, \theta_{R} ) e^{i\alpha}\psi
=
\int d^{4}x\; \overline{\psi} \psi 
=
\int d^{4}x\; \overline{\psi} \star \psi \,.
\ene
The above property is important for having a consistent gauge field theory on both commutative and noncommutative space.

%%%%%%%%%%%%%%%%%%%%%%%%%%%%%%%%%%%%
%\section{Dynamics of a Charged Fermion}\label{sec:Dynamics:Dirac}
\section{Local Gauge Fields Theory}\label{sec:LocalGaugeTheory}
In this section we will study local gauge field theory defined on the chiral spin noncommutative space that have been introduced in last section. Without loss of generality, we will assume that the matter field is spinor. 
\subsection{The Lagrangian for a free Dirac Spinor}

For a free spinor particle, chiral spin noncommutative extension of the Lagrangian density can be defined as
%\bee
%\widetilde{S} = \int \d x^{4} \widetilde{\call}_{0}
%\ene
%
\bee
\widetilde{\call}_{0} 
= 
\overline{\widetilde{\psi(x)}} \left( \gamma_{\mu}p^{\mu} - m \right)\widetilde{\psi(x)} \,.
\ene
Up to a surface term the above Lagrangian can be written in terms of ordinary fields as
\bee
\widetilde{\call}_{0} 
= 
\overline{\psi}(x) \overrightarrow{\calt}^{-1}
\left( \gamma_{\mu}p^{\mu} - m \right)
\overrightarrow{\calt}\psi(x) \,.
\ene
Because the translation operator $\overrightarrow{\calt}$ has non-trivial spinor structures, and hence $\overrightarrow{\calt}^{-1}\gamma_{\mu}\overrightarrow{\calt}\neq\gamma_{\mu}$. Therefore, if the $\widetilde{\call}_{0} $ is interpreted as the Lagrangian of a free spinor particle on the chiral spin noncommutative space, then the dispersion relation of a free particle receives corrections. In this paper, we introduce following definition as the Lagrangian of a free particle,
\bee
\widetilde{\call}_{0} 
= 
\overline{\widetilde{\psi(x)}} \left( \widetilde{\gamma}_{\mu}p^{\mu} - m \right)\widetilde{\psi(x)}\,,
\ene
where
\bee
\widetilde{\gamma}_{\mu}
=
\overrightarrow{\calt}\gamma_{\mu}\overrightarrow{\calt}^{-1}\,.
\ene
In this model, one can easily see that there is no noncommutative correction on a free spinor particle, \ie, $\widetilde{\call}_{0} = \call_{0} $.
%\bee
%\widetilde{\call}_{0} 
%= 
%\overline{\psi}(x) \left( \gamma_{\mu}p^{\mu} - m \right)\psi(x)
%\ene

\subsection{Local Gauge Transformation}
In commutative space, the local gauge transformation for the (spinor) matter field is given as $\psi'(x)=e^{i\alpha(x)}\psi(x)$. A direct noncommutative extension of the gauge transformation is $\widetilde{\psi(x)}'=e^{i\alpha(x)} \star \psi(x)=\left[ \overrightarrow{\calt}e^{i\alpha(x)} \right]\widetilde{\psi(x)}$. However, this kind of extension can not generally preserve the ordinary local gauge invariance which is important to have a consistent interpretations of the physical quantities on commutative and noncommutative space. In order to solve this problem, we propose following local gauge transformation for the matter field,
\bee
\widetilde{\psi(x)}'
= 
\widetilde{\psi'(x)}\,.
\ene
This means the local gauge transformation on noncommutative space is induced by a local gauge transformation on commutative space. The above transformation can be rewritten as
\bee
\widetilde{\psi(x)}'
= 
\widetilde{U}(x)\widetilde{\psi(x)} \,,
\ene
where the local gauge transformation operator $\widetilde{U}(x)$ is given as
\bee
\widetilde{U}(x)
=
\overrightarrow{\calt} e^{i\alpha(x)} \overrightarrow{\calt}^{-1} \,.
\ene
Apparently, the corresponding inverse gauge transformation $\overrightarrow{U}^{-1}(x) = \overrightarrow{\calt} e^{ -i\alpha(x)} \overrightarrow{\calt}^{-1} $. One can also easily obtain that $\overrightarrow{U}^{\dag}(x) = \overleftarrow{\calt}^{-1} e^{ -i\alpha(x)} \overleftarrow{\calt}$. Furthermore, up to total derivative terms, one has the relation $\overrightarrow{U}^{\dag}(x) = \overrightarrow{U}^{-1}(x)$, which means gauge transformation of the conjugate spinor can be written as,
$
\overline{ \widetilde{\psi(x)}' }
= 
\overline{ \widetilde{U}(x) \widetilde{ \psi(x)} } 
=
\overline{  \widetilde{ \psi(x)} } \overrightarrow{U}^{-1}(x) \,.
$
Hence, the total probability is preserved on both commutative and noncommutative space. 

Furthermore, since the noncommutative gauge potential $\widetilde{A^{\mu}(x)}$ has nontrivial spinor structure, when the minimal coupling $\gamma_{\mu}A^{\mu}$ in the original local gauge field theory, we have to define unambiguously the ordering between the noncommutative Dirac matrix $\widetilde{\gamma}_{\mu}$ and the noncommutative gauge potential $\widetilde{A^{\mu}(x)}$. A symmetric ordering, \ie, $\widetilde{\gamma}_{\mu}\widetilde{A^{\mu}(x)} + \widetilde{A^{\mu}(x)}\widetilde{\gamma}_{\mu}$ was proposed. However, the symmetric ordering is just a special case of a general ordering form. Here, we only require that extension of the minimal coupling satisfies the relativistic Hermitian. Explicitly, we define the full Lagrangian for a spinor particle coupling to gauge potential as follows,
\bee\label{eq:nc:Lag}
\widetilde{\call}
= 
\overline{\widetilde{\psi(x)}} \left[ 
\widetilde{\gamma}_{\mu} p^{\mu}  
- \frac{g}{2} \left( e^{i\kappa} \, \widetilde{\gamma}_{\mu} \widetilde{A^{\mu}(x)} 
+ e^{-i\kappa}\widetilde{A^{\mu}(x)} \widetilde{\gamma}_{\mu} \right)
\right]\widetilde{\psi(x)}
%\widetilde{\psi(x)}
%\overline{\widetilde{\psi(x)}} \widetilde{\gamma}_{\mu} \left( p^{\mu} - \widetilde{A^{\mu}(x)} \right)\widetilde{\psi(x)}
- m \overline{\widetilde{\psi(x)}} \widetilde{\psi(x)}\,,
\ene
where an exponential factor $e^{i\kappa}$ is introduced for generality, and without loss of generality $\kappa$ is assumed to be real constant. By using following relations,
\bea
\overrightarrow{U} \widetilde{\gamma}_{\mu} \overrightarrow{U}^{-1} 
&=&
\overrightarrow{\calt} e^{i\alpha(x)} \overrightarrow{\calt}^{-1}
\overrightarrow{\calt}\gamma_{\mu}\overrightarrow{\calt}^{-1}
\overrightarrow{\calt} e^{-i\alpha(x)} \overrightarrow{\calt}^{-1}
=
\widetilde{\gamma}_{\mu}\,,
\\[2mm]
\overrightarrow{U} p^{\mu} \overrightarrow{U}^{-1} 
&=&
p^{\mu} + i \overrightarrow{U}  \left[ \partial^{\mu} \overrightarrow{U}^{-1} \right]\,,
\ena
one can easily find that if the gauge transformation of the noncommutative gauge potential is defined as follows,
\bee\label{eq:GaugTrans:A}
\widetilde{A^{\mu}}'
=
\overrightarrow{U} \widetilde{A^{\mu}} \overrightarrow{U}^{-1}
-  \frac{ i }{ e } \overrightarrow{U}  \left[ \partial^{\mu} \overrightarrow{U}^{-1} \right] \,,
\ene
where $e=g\cos{\kappa}$ ($e=|e|$ (for electrodynamics)), then the Lagrangian is gauge invariant. Here and after we will use the normalization $e=g\cos{\kappa}$, and hence the allowed range of the parameter $\kappa$ is restricted in the region $(-\pi/2,\, \pi/2)$.

Up to a surface term, the noncommutative Lagrangian density $\widetilde{\call}$ can be rewritten in terms of commutative fields as follows,
\bee
\widetilde{\call}
= 
\overline{\psi(x)} \left[ 
\gamma_{\mu} p^{\mu}  
- \frac{g}{2} \left( e^{i\kappa} \gamma_{\mu} \cala^{\mu}(x)
+ e^{-i\kappa}\cala^{\mu}(x) \gamma_{\mu} \right)
\right] \psi(x)
- m \overline{\psi(x)} \psi(x)\,,
\ene
where\,,
%\bee
%\cala^{\mu}(x) = 
%\overrightarrow{\calt}^{-1} 
%\left[ \overrightarrow{\calt} A^{\mu}(x) \right] \overrightarrow{\calt} \,.
%\ene
%
\bea
\cala^{\mu}(x) 
&=& 
\left[ \overrightarrow{\calt} A^{\mu}(x) \right]
+ M^{\mu}(x)\overrightarrow{\calt} \,,
\\[3mm]
M^{\mu}(x) 
&=&
\left[ \overrightarrow{\calt}^{-1},\, \left[\overrightarrow{\calt} A^{\mu}(x)\right] \right] \,.
\ena
One can see clearly that there are two kinds of corrections; the first kind of correction is proportional to $\overrightarrow{\calt} A^{\mu}(x)$, and hence involves higher order derivatives of the original gauge potential $A^{\mu}$; while the second kind of correction involves higher order derivatives of both the original gauge potential $A^{\mu}$ and the matter field $\psi(x)$, since it is proportional to the term $M^{\mu}(x)\overrightarrow{\calt}$, and hence. Furthermore, the second kind of correction also contributes the Noether current, and leading order correction is given as
\bee
\widetilde{J}^{\mu}
=
- i \frac{ \delta \call }{ \delta( \partial_{\mu}\psi ) } \psi
=
J^{\mu} + \frac{i}{2} g \, \overline{\psi} 
\left( e^{i\kappa} \gamma^{\alpha} M_{\alpha} +  e^{-i\kappa} M_{\alpha} \gamma^{\alpha}  \right)
\gamma_{\nu} \big(\theta^{\mu\nu}_{L}\gamma_{L} + \theta^{\mu\nu}_{R}\gamma_{R} \big) \psi \,, 
\ene
where $J^{\mu}=\overline{\psi}\gamma^{\mu}\psi$ is the ordinary Noether current. Since the leading non-zero term of the function $M^{\mu}(x)$ is 
$M^{\alpha} 
\approx 
-\gamma_{\nu} \big(\theta^{\mu\nu}_{L}\gamma_{L} + \theta^{\mu\nu}_{R}\gamma_{R} \big) 
\partial_{\mu}A^{\alpha}
$
therefore, the non-trivial corrections on the Noether current starts from second order of the noncommutative parameters $\theta^{\mu\nu}_{L/R}$. 

For clarity, here we introduce noncommutative extensions of the ordinary covariant derivatives $D_{\mu}$. As we have explained, there is an ambiguity in the definition of noncommutative minimal coupling. Therefore, there are two kinds of noncommutative covariant derivatives $\widetilde{D}_{\mu}^{+}$ and $\widetilde{D}_{\mu}^{-}$ which are defined as follows,
\bea
\widetilde{D}_{\mu}^{+}
&=&
\partial_{\mu} + i g\,e^{i\kappa} \, \widetilde{A^{\mu}(x)} \,,
\\[2mm]
\widetilde{D}_{\mu}^{-}
&=&
\partial_{\mu} + i g\,e^{-i\kappa} \, \widetilde{A^{\mu}(x)} \,.
\ena
Then the Lagrangian density \eqref{eq:nc:Lag} can be rewritten as follows,
\bee
\widetilde{\call}
= \frac{1}{2} \;
\overline{\widetilde{\psi(x)}} \left[ 
i \,\widetilde{\gamma}^{\mu} \widetilde{D}_{\mu}^{+}
+ i \, \widetilde{D}_{\mu}^{-} \widetilde{\gamma}^{\mu}  
\right]\widetilde{\psi(x)}
- m \overline{\widetilde{\psi(x)}} \widetilde{\psi(x)}\,,
\ene

\subsection{Anomalous Magnetic and Electric Dipole Moments}
In contrast to the Noether current on which the leading order noncommutative correction is of second order of the noncommutative parameters, the leading order noncommutative corrections on the anomalous magnetic and electric dipole moments are linear of the noncommutative parameters. Ignoring the corrections involves derivatives of the matter field, the interaction Lagrangian is given as,
\bee
\widetilde{\call}_{I}
= 
- \frac{g}{2} \; \overline{\psi(x)} \left[ 
  e^{i\kappa} \gamma_{\mu} \left( \overrightarrow{\calt} A^{\mu}(x) \right)
+ e^{-i\kappa} \left( \overrightarrow{\calt} A^{\mu}(x)\right) \gamma_{\mu} 
\right] \psi(x)\,.
\ene
One can easily see that the $0$th order contribution is completely the same with ordinary minimal coupling, except for that the coupling $e$ is replaced by $g\cos{\kappa}$. 

On the other hand, the leading order non-trivial term of $\overrightarrow{\calt} A^{\mu}(x)$ is given as,
\bee
\overrightarrow{\calt} A^{\mu}(x)
\approx
\gamma_{\beta} \big(\theta^{\alpha\beta}_{L}\gamma_{L} + \theta^{\alpha\beta}_{R}\gamma_{R} \big)
\partial_{\alpha}A^{\mu}(x)\,.
\ene
Here we will study a simplified case in which $\theta^{\mu\nu}_{L/R} = \eta^{\mu\nu}\theta_{L/R}$. In this case, there are only three free parameters $\kappa$ and $\theta_{L/R}$. Inserting into the interaction Lagrangian we have,
\bee
\widetilde{\call}_{I}
= 
- e \, \overline{\psi(x)}\left[
\gamma_{\mu} A^{\mu}(x) 
+ \frac{a_{M}}{4m} \sigma_{\mu\nu} F^{\mu\nu} 
+ i \frac{a_{E}}{4m}\sigma_{\mu\nu} \gamma_{5}F^{\mu\nu} 
\right]  \psi(x)\,,
\ene
where the anomalous magnetic and electric dipole moments are given as
\bea
a_{M}
&=&
m\, \theta_{V} \tan{\kappa} \,,
\\
a_{E}
&=&
m\, \theta_{A}\,
\ena
and $\theta_{V}= \theta_{R} + \theta_{L}$ and $\theta_{A}= \theta_{R} - \theta_{L}$ are the vector and axial vector noncommutative parameters, respectively. Interestingly, those corrections depend on the mass of the matter particles. Here we consider the noncommutative corrections on muon-lepton. For the electric dipole moment the current world average is given as~\cite{Tanabashi:2018oca}
\bee
d^{\mu^{-}}_{E} = -0.1 \pm 0.9\times 10^{-21} \;e{\rm m}\,.
\ene
By requiring the noncommutative correction is consistent with experimental results in $1\sigma$ level, then we have
\bee\label{eq:bound:ae}
\big|\theta_{A}^{\mu^{-}} \big| \le 9.1\times 10^{-11}\, {\rm TeV}^{-1} \,.
\ene
On the other hand, the current experimental result for the anomalous magnetic dipole moment is given as~\cite{Tanabashi:2018oca},
\bee
a_{M}^{\mu^{-}} = (11659208.9 \pm 6.33)\times 10^{-10}\,.
\ene
By requiring the noncommutative contribution lies in the $1\sigma$ band (there is about $3.3\sigma$ discrepancy when comparing to the Standard Model prediction~\cite{Hagiwara:2017lse}, but here we don't consider this effect), then we have,
\bee
\theta_{V}^{\mu^{-}} \tan{\kappa^{\mu^{-}}}
\le 5.97 \times 10^{-6}\,{\rm TeV}^{-1} \,.
\ene
In case of the the minimal coupling is put in symmetric ordering, \ie, $\tan{\kappa^{\mu^{-}}}= 1$, then the vector noncommutative parameter is restricted seriously, $\theta_{V}^{\mu^{-}} 
\le 5.97 \times 10^{-6}\,{\rm TeV}^{-1}$. However, the symmetric ordering is not mandatary, therefore there are still parameter space that can be accessible by current experimental studies.  On the other hand, in parameter space with $\theta_{V}^{\mu^{-}} 
\sim1 \,{\rm TeV}^{-1}$,  the spin noncommutativity \eqref{eq:generalSpinNC} is nearly of vector form because strong bound on the axial noncommutative parameter (in this simplified model), see \eqref{eq:bound:ae}.

% For electron and muon, the corresponding corrections are
%\bea
%a_{M}^{e^{-}}
%&\approx&
%5.11 \times 10^{-7} \theta_{V} \tan{\kappa} \,,
%\\
%a_{E}^{e^{-}}
%&\approx&
%5.11 \times 10^{-7} \theta_{A} \,,
%\\
%a_{M}^{\mu^{-}}
%&\approx&
%1.06 \times 10^{-4} \theta_{V} \tan{\kappa} \,,
%\\
%a_{E}^{\mu^{-}}
%&\approx&
%1.06 \times 10^{-4} \theta_{A} \,,
%\ena
%where the noncommutative parameters are expressed in unit of ${\rm TeV}^{-1}$.

\subsection{Equation of Motion of Electromagnetic Field}\label{sec:Dynamics:Photon}
In this subsection we study the equation of motion of the gauge field $\widetilde{A}_{\mu}(x)$. The gauge transformation of the gauge field $\widetilde{A}_{\mu}(x)$ has been given in \eqref{eq:GaugTrans:A}. It is apparently that the gauge field $\widetilde{A}_{\mu}(x)$ behaves like a non-Abel gauge field, hence the corresponding field strength tensor can be defined as the usually. However, since there are two kinds of noncommutative covariant derivatives, a proper definition of the field strength tensor is given as follows,
\bee
\widetilde{F}_{\mu\nu}
=
\frac{i}{2e}
\left\{ \left[  \widetilde{D}_{\mu}^{+},\,  \widetilde{D}_{\nu}^{-}\right] 
+
\left[  \widetilde{D}_{\mu}^{-},\,  \widetilde{D}_{\nu}^{+}\right] \right\}
= 
\left[ \overrightarrow{\calt} F_{\mu\nu} \right]
- i e\left[ \widetilde{A}_{\mu},\, \widetilde{A}_{\nu}\right] \,.
\ene
Compared to the usual field strength tensor $F_{\mu\nu}(x)$, there are two kinds of corrections. The first kind of correction is characterized by a direct action of the translation operator $\overrightarrow{\calt}$ on the ordinary field strength tensor $F_{\mu\nu}(x)$, while the second kind of correction is induced by the non-Abel property of the gauge field $\widetilde{A}_{\mu}(x)$, and is of course proportional to the gauge coupling constant $g$. It is also precisely because of this, up to leading order of the noncommutative parameter $\theta^{\mu\nu}_{L/R}$, only the first kind of correction is non-trivial and given as follows,
\bee
\widetilde{F}_{\mu\nu}
\approx
F_{\mu\nu}
+
\gamma_{\beta} \big(\theta^{\alpha\beta}_{L}\gamma_{L} + \theta^{\alpha\beta}_{R}\gamma_{R} \big) \partial_{\alpha}F_{\mu\nu} \,.
\ene
On the other hand, the leading non-trivial corrections due to non-Abel property is given as
\bee
\widetilde{F_{I}}_{\mu\nu}
=
g \left[
 i  \theta_{+}^{\alpha\beta\lambda\delta} \sigma_{\lambda\delta} 
- \gamma_{5} \theta_{-}^{\alpha\beta}\right]
\left[\partial_{\alpha} A_{\mu} \right]
\left[\partial_{\beta} A_{\nu} \right] \,,
\ene
%\bee
%\widetilde{F}_{\mu\nu}
%=
%F_{\mu\nu}
%+
%\gamma_{\beta} \big(\theta^{\alpha\beta}_{L}\gamma_{L} + \theta^{\alpha\beta}_{R}\gamma_{R} \big) \partial_{\alpha}F_{\mu\nu}
%+ 
%g \left[
%- i  \theta_{+}^{\alpha\beta\lambda\delta} \sigma_{\lambda\delta} 
%+ \gamma_{5} \theta_{-}^{\alpha\beta}\right]
%\left[\partial_{\alpha} A_{\mu} \right]
%\left[\partial_{\beta} A_{\nu} \right]
%\ene
where the tensors $\theta_{+}^{\mu\nu\lambda\delta}$ and $\theta_{-}^{\mu\nu}$ are given as,
\bea
\theta_{+}^{\mu\nu\lambda\delta} 
&=& \big( \theta^{\mu\lambda}_{L}\theta^{\nu\delta}_{R} + \theta^{\mu\lambda}_{R}\theta^{\nu\delta}_{L}  \big) \,,
\\[2mm]
\theta_{-}^{\mu\nu} 
&=&
\big( \theta^{\mu\delta}_{L}\theta^{\nu}_{R\delta} -\theta^{\mu\delta}_{R}\theta^{\nu}_{L\delta} \big)\,.  
\ena

On the other hand, since the field strength tensor has non-trivial spinor structure, the corresponding Lagrangian density is defined as,
\bee
\widetilde{\call}_{G} 
=
- \frac{1}{16} {\rm Tr}\left\{ \widetilde{F}_{\mu\nu} \widetilde{F}^{\mu\nu} \right\} \,,
\ene
where the trace $``{\rm Tr}"$ is for the spinor degree of freedom. Up to leading order noncommutative correction, the Lagrangian density is given as,
\bee
\widetilde{\call}_{G} 
\approx
- \frac{1}{4} F_{\mu\nu} F^{\mu\nu} 
- \frac{1}{4}\theta_{+}^{\alpha\beta} 
\left[ \partial_{\alpha}F_{\mu\nu} \right] \left[ \partial_{\beta}F^{\mu\nu} \right]  \,,
\ene
where the constant tensor $\theta_{+}^{\alpha\beta}$ is given as,
\bee
\theta_{+}^{\alpha\beta} 
=
\big( \theta^{\alpha\mu}_{L}\theta^{\beta}_{R\mu} + \theta^{\alpha\nu}_{R}\theta^{\beta}_{L\nu}  \big) \,. 
\ene
The corresponding equation of motion can be easily obtained,
\bee
\left( 1 - \theta_{+}^{\alpha\beta} \partial_{\alpha}\partial_{\beta} \right) \Box A_{\mu} =0 \,.
\ene
One can see that for a free ordinary gauge potential $A_{\mu}$, it also satisfies above equation. Therefore, up to second order of the noncommutative parameters, there is no net correction. However, one can also easily see that noncommutative corrections can be nontrivial at higher orders of noncommutative parameters, particularly the non-Abel corrections, and we will study those effects in the future.

%\bee
%\widetilde{\call}_{G} 
%=
%- \frac{1}{4} {\rm Tr}\left\{ \left[ \overrightarrow{\calt}F_{\mu\nu} \right] \left[ \overrightarrow{\calt}F_{\mu\nu} \right] \right\}
%\ene

%\bee
%\left[ \widetilde{A}_{\alpha}(x), \widetilde{A}_{\beta}(x) \right]
%\approx \left[
%- i  \theta_{+}^{\mu\nu\lambda\delta} \sigma_{\lambda\delta} 
%+ \gamma_{5} \theta_{-}^{\mu\nu}\right]
%\left[\partial_{\mu}A_{\alpha}(x) \right]
%\left[\partial_{\nu}A_{\beta}(x) \right]
%\ene

\section{Summary and Conclusions}\label{sec:sum}
In summary we introduced a class of spin noncommutative space, and particularly the coordinate operators are assumed to be chirality dependent. As usual, noncommutative extensions of the classical fields are defined via Weyl ordering. We find that noncommutative effect is trivial in the integral of product of two noncommutative functions. This property keeps the total probability invariant, which is important to have a consistent extension of the probability interpretation in ordinary Quantum Mechanics (QM). 

Furthermore, we constructed a noncommutative extension of the local gauge field theory. Without loss of generality, taking spinor matter field as an example, gauge transformations of the matter and gauge fields are unambiguously defined. In our model, the noncommutative extensions of the gauge transformations are induced by the ordinary gauge transformations, and hence there is no any ambiguity in the physical interpretations of the fields. However, since our noncommutative model is spin-dependent, the there is an ambiguity in extension of the ordinary minimal coupling, \ie, the ordering between the Dirac matrix $\gamma_{\mu}$ and the noncommutative gauge potential $\widetilde{A}^{\mu}$. In contrast to the usual way, in which a symmetric ordering is adopted, we introduce a more general parameterization: phase factors $e^{i\kappa}$ and its conjugate were introduced for two kinds of ordering to resolve the above ambiguity.

We studied equations of motion of the matter and gauge fields. We find that, up to leading order of the noncommutative parameters, while noncommutative corrections on the gauge potential is trivial, there are two kinds of corrections on the matter field. The first kind of correction involves derivatives of the matter field, and hence a nontrivial contribution to the ordinary Noether current. The second kind of corrections depends only on derivatives of the gauge potential, and an immediate contribution to the anomalous magnetic and electric dipole moments. Furthermore, in our normalization of the gauge coupling constant, we find that while the contribution to anomalous electric dipole moment depends on the axial noncommutative parameter $\theta_{A}$, corrections on the anomalous magnetic dipole moment depends not only the vector noncommutative parameter $\theta_{V}$ but also the ordering parameter $\kappa$. Moreover, taking the muon lepton as an example, we studied the current experimental constrains on the noncommutative parameters in case of $\theta_{L/R}^{\mu\nu}=\theta_{L/R}\eta^{\mu\nu}$. We find that in the parameter space region where $\theta_{V}$ is accessible by current experiments ($\theta_{V} \sim 1{\rm TeV}^{-1}$), the chirality effect of this simplified model is negligible, \ie, $\theta_{L}\sim \theta_{R}$, meanwhile ordering between $\gamma_{\mu}$ and $\widetilde{A}^{\mu}$ is required to be heavily sloped, $\kappa \sim 10^{-6}$. While this effect seems un-natural, there are still lost of interesting parameters space in more general models which will be studied in future.

%%%%%%%%%%%%%%%
\section*{Acknowledgements}
K. M. is supported by the National Natural Science Foundation of China under Grant No.11705113, and the Natural Science Basic Research Plan in Shaanxi Province of China under Grant No. 2018JQ1018, as well as the Scientific Research Program Funded by Shaanxi Provincial Education Department under Grant No. 18JK0153.

\bibliography{aString}

\begin{thebibliography}{52}%
\makeatletter
\providecommand \@ifxundefined [1]{%
 \@ifx{#1\undefined}
}%
\providecommand \@ifnum [1]{%
 \ifnum #1\expandafter \@firstoftwo
 \else \expandafter \@secondoftwo
 \fi
}%
\providecommand \@ifx [1]{%
 \ifx #1\expandafter \@firstoftwo
 \else \expandafter \@secondoftwo
 \fi
}%
\providecommand \natexlab [1]{#1}%
\providecommand \enquote  [1]{``#1''}%
\providecommand \bibnamefont  [1]{#1}%
\providecommand \bibfnamefont [1]{#1}%
\providecommand \citenamefont [1]{#1}%
\providecommand \href@noop [0]{\@secondoftwo}%
\providecommand \href [0]{\begingroup \@sanitize@url \@href}%
\providecommand \@href[1]{\@@startlink{#1}\@@href}%
\providecommand \@@href[1]{\endgroup#1\@@endlink}%
\providecommand \@sanitize@url [0]{\catcode `\\12\catcode `\$12\catcode
  `\&12\catcode `\#12\catcode `\^12\catcode `\_12\catcode `\%12\relax}%
\providecommand \@@startlink[1]{}%
\providecommand \@@endlink[0]{}%
\providecommand \url  [0]{\begingroup\@sanitize@url \@url }%
\providecommand \@url [1]{\endgroup\@href {#1}{\urlprefix }}%
\providecommand \urlprefix  [0]{URL }%
\providecommand \Eprint [0]{\href }%
\providecommand \doibase [0]{http://dx.doi.org/}%
\providecommand \selectlanguage [0]{\@gobble}%
\providecommand \bibinfo  [0]{\@secondoftwo}%
\providecommand \bibfield  [0]{\@secondoftwo}%
\providecommand \translation [1]{[#1]}%
\providecommand \BibitemOpen [0]{}%
\providecommand \bibitemStop [0]{}%
\providecommand \bibitemNoStop [0]{.\EOS\space}%
\providecommand \EOS [0]{\spacefactor3000\relax}%
\providecommand \BibitemShut  [1]{\csname bibitem#1\endcsname}%
\let\auto@bib@innerbib\@empty
%</preamble>
\bibitem [{\citenamefont {Freidel}\ and\ \citenamefont
  {Livine}(2006)}]{Freidel:2005me}%
  \BibitemOpen
  \bibfield  {author} {\bibinfo {author} {\bibfnamefont {Laurent}\ \bibnamefont
  {Freidel}}\ and\ \bibinfo {author} {\bibfnamefont {Etera~R.}\ \bibnamefont
  {Livine}},\ }\bibfield  {title} {\enquote {\bibinfo {title} {{Effective 3-D
  quantum gravity and non-commutative quantum field theory}},}\ }\bibfield
  {booktitle} {\emph {\bibinfo {booktitle} {{4th International Symposium on
  Quantum Theory and Symmetries and 6th International Workshop on Lie Theory
  and Its Applications in Physics (QTS-4) (LT-6) Varna, Bulgaria, August 15-21,
  2005}}},\ }\href {\doibase 10.1103/PhysRevLett.96.221301} {\bibfield
  {journal} {\bibinfo  {journal} {Phys. Rev. Lett.}\ }\textbf {\bibinfo
  {volume} {96}},\ \bibinfo {pages} {221301} (\bibinfo {year} {2006})},\
  \Eprint {http://arxiv.org/abs/hep-th/0512113} {arXiv:hep-th/0512113 [hep-th]}
  \BibitemShut {NoStop}%
%%CITATION = HEP-TH/0512113;%%
\bibitem [{\citenamefont {Moffat}(2000{\natexlab{a}})}]{Moffat:2000gr}%
  \BibitemOpen
  \bibfield  {author} {\bibinfo {author} {\bibfnamefont {J.~W.}\ \bibnamefont
  {Moffat}},\ }\bibfield  {title} {\enquote {\bibinfo {title} {{Noncommutative
  quantum gravity}},}\ }\href {\doibase 10.1016/S0370-2693(00)01064-9}
  {\bibfield  {journal} {\bibinfo  {journal} {Phys. Lett.}\ }\textbf {\bibinfo
  {volume} {B491}},\ \bibinfo {pages} {345--352} (\bibinfo {year}
  {2000}{\natexlab{a}})},\ \Eprint {http://arxiv.org/abs/hep-th/0007181}
  {arXiv:hep-th/0007181 [hep-th]} \BibitemShut {NoStop}%
%%CITATION = HEP-TH/0007181;%%
\bibitem [{\citenamefont {Moffat}(2000{\natexlab{b}})}]{Moffat:2000fv}%
  \BibitemOpen
  \bibfield  {author} {\bibinfo {author} {\bibfnamefont {J.~W.}\ \bibnamefont
  {Moffat}},\ }\bibfield  {title} {\enquote {\bibinfo {title} {{Perturbative
  noncommutative quantum gravity}},}\ }\href {\doibase
  10.1016/S0370-2693(00)01139-4} {\bibfield  {journal} {\bibinfo  {journal}
  {Phys. Lett.}\ }\textbf {\bibinfo {volume} {B493}},\ \bibinfo {pages}
  {142--148} (\bibinfo {year} {2000}{\natexlab{b}})},\ \Eprint
  {http://arxiv.org/abs/hep-th/0008089} {arXiv:hep-th/0008089 [hep-th]}
  \BibitemShut {NoStop}%
%%CITATION = HEP-TH/0008089;%%
\bibitem [{\citenamefont {Faizal}(2013)}]{Faizal:2013ioa}%
  \BibitemOpen
  \bibfield  {author} {\bibinfo {author} {\bibfnamefont {Mir}\ \bibnamefont
  {Faizal}},\ }\bibfield  {title} {\enquote {\bibinfo {title} {{Noncommutative
  Quantum Gravity}},}\ }\href {\doibase 10.1142/S021773231350034X} {\bibfield
  {journal} {\bibinfo  {journal} {Mod. Phys. Lett.}\ }\textbf {\bibinfo
  {volume} {A28}},\ \bibinfo {pages} {1350034} (\bibinfo {year} {2013})},\
  \Eprint {http://arxiv.org/abs/1302.5156} {arXiv:1302.5156 [gr-qc]}
  \BibitemShut {NoStop}%
%%CITATION = ARXIV:1302.5156;%%
\bibitem [{\citenamefont {Seiberg}\ and\ \citenamefont
  {Witten}(1999)}]{Seiberg:1999vs}%
  \BibitemOpen
  \bibfield  {author} {\bibinfo {author} {\bibfnamefont {Nathan}\ \bibnamefont
  {Seiberg}}\ and\ \bibinfo {author} {\bibfnamefont {Edward}\ \bibnamefont
  {Witten}},\ }\bibfield  {title} {\enquote {\bibinfo {title} {{String theory
  and noncommutative geometry}},}\ }\href {\doibase
  10.1088/1126-6708/1999/09/032} {\bibfield  {journal} {\bibinfo  {journal}
  {JHEP}\ }\textbf {\bibinfo {volume} {09}},\ \bibinfo {pages} {032} (\bibinfo
  {year} {1999})},\ \Eprint {http://arxiv.org/abs/hep-th/9908142}
  {arXiv:hep-th/9908142 [hep-th]} \BibitemShut {NoStop}%
%%CITATION = HEP-TH/9908142;%%
\bibitem [{\citenamefont {Gamboa}\ \emph {et~al.}(2001)\citenamefont {Gamboa},
  \citenamefont {Loewe},\ and\ \citenamefont {Rojas}}]{Gamboa:2000yq}%
  \BibitemOpen
  \bibfield  {author} {\bibinfo {author} {\bibfnamefont {J.}~\bibnamefont
  {Gamboa}}, \bibinfo {author} {\bibfnamefont {M.}~\bibnamefont {Loewe}}, \
  and\ \bibinfo {author} {\bibfnamefont {J.~C.}\ \bibnamefont {Rojas}},\
  }\bibfield  {title} {\enquote {\bibinfo {title} {{Noncommutative quantum
  mechanics}},}\ }\href {\doibase 10.1103/PhysRevD.64.067901} {\bibfield
  {journal} {\bibinfo  {journal} {Phys. Rev.}\ }\textbf {\bibinfo {volume}
  {D64}},\ \bibinfo {pages} {067901} (\bibinfo {year} {2001})},\ \Eprint
  {http://arxiv.org/abs/hep-th/0010220} {arXiv:hep-th/0010220 [hep-th]}
  \BibitemShut {NoStop}%
%%CITATION = HEP-TH/0010220;%%
\bibitem [{\citenamefont {Ho}\ and\ \citenamefont {Kao}(2002)}]{Ho:2001aa}%
  \BibitemOpen
  \bibfield  {author} {\bibinfo {author} {\bibfnamefont {Pei-Ming}\
  \bibnamefont {Ho}}\ and\ \bibinfo {author} {\bibfnamefont {Hsien-Chung}\
  \bibnamefont {Kao}},\ }\bibfield  {title} {\enquote {\bibinfo {title}
  {{Noncommutative quantum mechanics from noncommutative quantum field
  theory}},}\ }\href {\doibase 10.1103/PhysRevLett.88.151602} {\bibfield
  {journal} {\bibinfo  {journal} {Phys. Rev. Lett.}\ }\textbf {\bibinfo
  {volume} {88}},\ \bibinfo {pages} {151602} (\bibinfo {year} {2002})},\
  \Eprint {http://arxiv.org/abs/hep-th/0110191} {arXiv:hep-th/0110191 [hep-th]}
  \BibitemShut {NoStop}%
%%CITATION = HEP-TH/0110191;%%
\bibitem [{\citenamefont {Douglas}\ and\ \citenamefont
  {Nekrasov}(2001)}]{Douglas:2001ba}%
  \BibitemOpen
  \bibfield  {author} {\bibinfo {author} {\bibfnamefont {Michael~R.}\
  \bibnamefont {Douglas}}\ and\ \bibinfo {author} {\bibfnamefont {Nikita~A.}\
  \bibnamefont {Nekrasov}},\ }\bibfield  {title} {\enquote {\bibinfo {title}
  {{Noncommutative field theory}},}\ }\href {\doibase
  10.1103/RevModPhys.73.977} {\bibfield  {journal} {\bibinfo  {journal} {Rev.
  Mod. Phys.}\ }\textbf {\bibinfo {volume} {73}},\ \bibinfo {pages} {977--1029}
  (\bibinfo {year} {2001})},\ \Eprint {http://arxiv.org/abs/hep-th/0106048}
  {arXiv:hep-th/0106048 [hep-th]} \BibitemShut {NoStop}%
%%CITATION = HEP-TH/0106048;%%
\bibitem [{\citenamefont {Szabo}(2003)}]{Szabo:2001kg}%
  \BibitemOpen
  \bibfield  {author} {\bibinfo {author} {\bibfnamefont {Richard~J.}\
  \bibnamefont {Szabo}},\ }\bibfield  {title} {\enquote {\bibinfo {title}
  {{Quantum field theory on noncommutative spaces}},}\ }\bibfield  {booktitle}
  {\emph {\bibinfo {booktitle} {{Frontiers of Mathematical Physics: Summer
  Workshop on Particles, Fields and Strings Burnaby, Canada, July 16-27,
  2001}}},\ }\href {\doibase 10.1016/S0370-1573(03)00059-0} {\bibfield
  {journal} {\bibinfo  {journal} {Phys. Rept.}\ }\textbf {\bibinfo {volume}
  {378}},\ \bibinfo {pages} {207--299} (\bibinfo {year} {2003})},\ \Eprint
  {http://arxiv.org/abs/hep-th/0109162} {arXiv:hep-th/0109162 [hep-th]}
  \BibitemShut {NoStop}%
%%CITATION = HEP-TH/0109162;%%
\bibitem [{\citenamefont {Ma}\ \emph {et~al.}(2018)\citenamefont {Ma},
  \citenamefont {Ren},\ and\ \citenamefont {Wang}}]{Ma:2017fnt}%
  \BibitemOpen
  \bibfield  {author} {\bibinfo {author} {\bibfnamefont {Kai}\ \bibnamefont
  {Ma}}, \bibinfo {author} {\bibfnamefont {Ya-Jie}\ \bibnamefont {Ren}}, \ and\
  \bibinfo {author} {\bibfnamefont {Ya-Hui}\ \bibnamefont {Wang}},\ }\bibfield
  {title} {\enquote {\bibinfo {title} {{Probing Noncommutativities of Phase
  Space by Using Persistent Charged Current and Its Asymmetry}},}\ }\href
  {\doibase 10.1103/PhysRevD.97.115011} {\bibfield  {journal} {\bibinfo
  {journal} {Phys. Rev.}\ }\textbf {\bibinfo {volume} {D97}},\ \bibinfo {pages}
  {115011} (\bibinfo {year} {2018})},\ \Eprint
  {http://arxiv.org/abs/1703.10923} {arXiv:1703.10923 [hep-th]} \BibitemShut
  {NoStop}%
%%CITATION = ARXIV:1703.10923;%%
\bibitem [{\citenamefont {Chaichian}\ \emph
  {et~al.}(2001{\natexlab{a}})\citenamefont {Chaichian}, \citenamefont
  {Demichev}, \citenamefont {Presnajder}, \citenamefont {Sheikh-Jabbari},\ and\
  \citenamefont {Tureanu}}]{Chaichian:2001pw}%
  \BibitemOpen
  \bibfield  {author} {\bibinfo {author} {\bibfnamefont {M.}~\bibnamefont
  {Chaichian}}, \bibinfo {author} {\bibfnamefont {A.}~\bibnamefont {Demichev}},
  \bibinfo {author} {\bibfnamefont {P.}~\bibnamefont {Presnajder}}, \bibinfo
  {author} {\bibfnamefont {M.~M.}\ \bibnamefont {Sheikh-Jabbari}}, \ and\
  \bibinfo {author} {\bibfnamefont {A.}~\bibnamefont {Tureanu}},\ }\bibfield
  {title} {\enquote {\bibinfo {title} {{Quantum theories on noncommutative
  spaces with nontrivial topology: Aharonov-Bohm and Casimir effects}},}\
  }\href {\doibase 10.1016/S0550-3213(01)00348-0} {\bibfield  {journal}
  {\bibinfo  {journal} {Nucl. Phys.}\ }\textbf {\bibinfo {volume} {B611}},\
  \bibinfo {pages} {383--402} (\bibinfo {year} {2001}{\natexlab{a}})},\ \Eprint
  {http://arxiv.org/abs/hep-th/0101209} {arXiv:hep-th/0101209 [hep-th]}
  \BibitemShut {NoStop}%
%%CITATION = HEP-TH/0101209;%%
\bibitem [{\citenamefont {Chaichian}\ \emph {et~al.}(2002)\citenamefont
  {Chaichian}, \citenamefont {Demichev}, \citenamefont {Presnajder},
  \citenamefont {Sheikh-Jabbari},\ and\ \citenamefont
  {Tureanu}}]{Chaichian:2000hy}%
  \BibitemOpen
  \bibfield  {author} {\bibinfo {author} {\bibfnamefont {M.}~\bibnamefont
  {Chaichian}}, \bibinfo {author} {\bibfnamefont {A.}~\bibnamefont {Demichev}},
  \bibinfo {author} {\bibfnamefont {P.}~\bibnamefont {Presnajder}}, \bibinfo
  {author} {\bibfnamefont {M.~M.}\ \bibnamefont {Sheikh-Jabbari}}, \ and\
  \bibinfo {author} {\bibfnamefont {A.}~\bibnamefont {Tureanu}},\ }\bibfield
  {title} {\enquote {\bibinfo {title} {{Aharonov-Bohm effect in noncommutative
  spaces}},}\ }\href {\doibase 10.1016/S0370-2693(02)01176-0} {\bibfield
  {journal} {\bibinfo  {journal} {Phys. Lett.}\ }\textbf {\bibinfo {volume}
  {B527}},\ \bibinfo {pages} {149--154} (\bibinfo {year} {2002})},\ \Eprint
  {http://arxiv.org/abs/hep-th/0012175} {arXiv:hep-th/0012175 [hep-th]}
  \BibitemShut {NoStop}%
%%CITATION = HEP-TH/0012175;%%
\bibitem [{\citenamefont {Li}\ and\ \citenamefont {Dulat}(2006)}]{Li:2005mi}%
  \BibitemOpen
  \bibfield  {author} {\bibinfo {author} {\bibfnamefont {Kang}\ \bibnamefont
  {Li}}\ and\ \bibinfo {author} {\bibfnamefont {Sayipjamal}\ \bibnamefont
  {Dulat}},\ }\bibfield  {title} {\enquote {\bibinfo {title} {{The
  Aharonov-Bohm effect in noncommutative quantum mechanics}},}\ }\href
  {\doibase 10.1140/epjc/s2006-02538-2} {\bibfield  {journal} {\bibinfo
  {journal} {Eur. Phys. J.}\ }\textbf {\bibinfo {volume} {C46}},\ \bibinfo
  {pages} {825--828} (\bibinfo {year} {2006})},\ \Eprint
  {http://arxiv.org/abs/hep-th/0508193} {arXiv:hep-th/0508193 [hep-th]}
  \BibitemShut {NoStop}%
%%CITATION = HEP-TH/0508193;%%
\bibitem [{\citenamefont {Chaichian}\ \emph {et~al.}(2008)\citenamefont
  {Chaichian}, \citenamefont {Langvik}, \citenamefont {Sasaki},\ and\
  \citenamefont {Tureanu}}]{Chaichian:2008gf}%
  \BibitemOpen
  \bibfield  {author} {\bibinfo {author} {\bibfnamefont {Masud}\ \bibnamefont
  {Chaichian}}, \bibinfo {author} {\bibfnamefont {Miklos}\ \bibnamefont
  {Langvik}}, \bibinfo {author} {\bibfnamefont {Shin}\ \bibnamefont {Sasaki}},
  \ and\ \bibinfo {author} {\bibfnamefont {Anca}\ \bibnamefont {Tureanu}},\
  }\bibfield  {title} {\enquote {\bibinfo {title} {{Gauge Covariance of the
  Aharonov-Bohm Phase in Noncommutative Quantum Mechanics}},}\ }\href {\doibase
  10.1016/j.physletb.2008.06.050} {\bibfield  {journal} {\bibinfo  {journal}
  {Phys. Lett.}\ }\textbf {\bibinfo {volume} {B666}},\ \bibinfo {pages}
  {199--204} (\bibinfo {year} {2008})},\ \Eprint
  {http://arxiv.org/abs/0804.3565} {arXiv:0804.3565 [hep-th]} \BibitemShut
  {NoStop}%
%%CITATION = ARXIV:0804.3565;%%
\bibitem [{\citenamefont {Chaichian}\ \emph
  {et~al.}(2001{\natexlab{b}})\citenamefont {Chaichian}, \citenamefont
  {Sheikh-Jabbari},\ and\ \citenamefont {Tureanu}}]{Chaichian:2000si}%
  \BibitemOpen
  \bibfield  {author} {\bibinfo {author} {\bibfnamefont {M.}~\bibnamefont
  {Chaichian}}, \bibinfo {author} {\bibfnamefont {M.~M.}\ \bibnamefont
  {Sheikh-Jabbari}}, \ and\ \bibinfo {author} {\bibfnamefont {A.}~\bibnamefont
  {Tureanu}},\ }\bibfield  {title} {\enquote {\bibinfo {title} {{Hydrogen atom
  spectrum and the Lamb shift in noncommutative QED}},}\ }\href {\doibase
  10.1103/PhysRevLett.86.2716} {\bibfield  {journal} {\bibinfo  {journal}
  {Phys. Rev. Lett.}\ }\textbf {\bibinfo {volume} {86}},\ \bibinfo {pages}
  {2716} (\bibinfo {year} {2001}{\natexlab{b}})},\ \Eprint
  {http://arxiv.org/abs/hep-th/0010175} {arXiv:hep-th/0010175 [hep-th]}
  \BibitemShut {NoStop}%
%%CITATION = HEP-TH/0010175;%%
\bibitem [{\citenamefont {Zhang}(2004)}]{Zhang:2004yu}%
  \BibitemOpen
  \bibfield  {author} {\bibinfo {author} {\bibfnamefont {Jian-zu}\ \bibnamefont
  {Zhang}},\ }\bibfield  {title} {\enquote {\bibinfo {title} {{Testing spatial
  noncommutativity via Rydberg atoms}},}\ }\href {\doibase
  10.1103/PhysRevLett.93.043002} {\bibfield  {journal} {\bibinfo  {journal}
  {Phys. Rev. Lett.}\ }\textbf {\bibinfo {volume} {93}},\ \bibinfo {pages}
  {043002} (\bibinfo {year} {2004})},\ \Eprint
  {http://arxiv.org/abs/hep-ph/0405143} {arXiv:hep-ph/0405143 [hep-ph]}
  \BibitemShut {NoStop}%
%%CITATION = HEP-PH/0405143;%%
\bibitem [{\citenamefont {Zhang}\ and\ \citenamefont
  {Horvathy}(2012{\natexlab{a}})}]{Zhang:2011kr}%
  \BibitemOpen
  \bibfield  {author} {\bibinfo {author} {\bibfnamefont {P.~M.}\ \bibnamefont
  {Zhang}}\ and\ \bibinfo {author} {\bibfnamefont {P.~A.}\ \bibnamefont
  {Horvathy}},\ }\bibfield  {title} {\enquote {\bibinfo {title} {{Kohn
  condition and exotic Newton-Hooke symmetry in the non-commutative Landau
  problem}},}\ }\href {\doibase 10.1016/j.physletb.2011.11.035} {\bibfield
  {journal} {\bibinfo  {journal} {Phys. Lett.}\ }\textbf {\bibinfo {volume}
  {B706}},\ \bibinfo {pages} {442--446} (\bibinfo {year}
  {2012}{\natexlab{a}})},\ \Eprint {http://arxiv.org/abs/1111.1595}
  {arXiv:1111.1595 [hep-th]} \BibitemShut {NoStop}%
%%CITATION = ARXIV:1111.1595;%%
\bibitem [{\citenamefont {Zhang}\ and\ \citenamefont
  {Horvathy}(2012{\natexlab{b}})}]{Zhang:2011zua}%
  \BibitemOpen
  \bibfield  {author} {\bibinfo {author} {\bibfnamefont {P-M.}\ \bibnamefont
  {Zhang}}\ and\ \bibinfo {author} {\bibfnamefont {P.~A.}\ \bibnamefont
  {Horvathy}},\ }\bibfield  {title} {\enquote {\bibinfo {title} {{Chiral
  Decomposition in the Non-Commutative Landau Problem}},}\ }\href {\doibase
  10.1016/j.aop.2012.02.014} {\bibfield  {journal} {\bibinfo  {journal} {Annals
  Phys.}\ }\textbf {\bibinfo {volume} {327}},\ \bibinfo {pages} {1730--1743}
  (\bibinfo {year} {2012}{\natexlab{b}})},\ \Eprint
  {http://arxiv.org/abs/1112.0409} {arXiv:1112.0409 [hep-th]} \BibitemShut
  {NoStop}%
%%CITATION = ARXIV:1112.0409;%%
\bibitem [{\citenamefont {Ma}(2017)}]{Ma:2017rwg}%
  \BibitemOpen
  \bibfield  {author} {\bibinfo {author} {\bibfnamefont {Kai}\ \bibnamefont
  {Ma}},\ }\bibfield  {title} {\enquote {\bibinfo {title} {{Constrains of
  Charge-to-Mass Ratios on Noncommutative Phase Space}},}\ }\href {\doibase
  10.1155/2017/1945156} {\bibfield  {journal} {\bibinfo  {journal} {Adv. High
  Energy Phys.}\ }\textbf {\bibinfo {volume} {2017}},\ \bibinfo {pages}
  {1945156} (\bibinfo {year} {2017})},\ \Eprint
  {http://arxiv.org/abs/1705.05789} {arXiv:1705.05789 [hep-ph]} \BibitemShut
  {NoStop}%
%%CITATION = ARXIV:1705.05789;%%
\bibitem [{\citenamefont {Ramírez}\ and\ \citenamefont
  {Deriglazov}(2017)}]{Ramirez:2017pmp}%
  \BibitemOpen
  \bibfield  {author} {\bibinfo {author} {\bibfnamefont {Walberto~Guzmán}\
  \bibnamefont {Ramírez}}\ and\ \bibinfo {author} {\bibfnamefont {Alexei~A.}\
  \bibnamefont {Deriglazov}},\ }\bibfield  {title} {\enquote {\bibinfo {title}
  {{Relativistic effects due to gravimagnetic moment of a rotating body}},}\
  }\href {\doibase 10.1103/PhysRevD.96.124013} {\bibfield  {journal} {\bibinfo
  {journal} {Phys. Rev.}\ }\textbf {\bibinfo {volume} {D96}},\ \bibinfo {pages}
  {124013} (\bibinfo {year} {2017})},\ \Eprint
  {http://arxiv.org/abs/1709.06894} {arXiv:1709.06894 [gr-qc]} \BibitemShut
  {NoStop}%
%%CITATION = ARXIV:1709.06894;%%
\bibitem [{\citenamefont {Ma}\ \emph {et~al.}(2016{\natexlab{a}})\citenamefont
  {Ma}, \citenamefont {Wang},\ and\ \citenamefont {Yang}}]{Ma:2016rhk}%
  \BibitemOpen
  \bibfield  {author} {\bibinfo {author} {\bibfnamefont {Kai}\ \bibnamefont
  {Ma}}, \bibinfo {author} {\bibfnamefont {Jian-Hua}\ \bibnamefont {Wang}}, \
  and\ \bibinfo {author} {\bibfnamefont {Huan-Xiong}\ \bibnamefont {Yang}},\
  }\bibfield  {title} {\enquote {\bibinfo {title} {{Time-dependent
  Aharonov–Bohm effect on the noncommutative space}},}\ }\href {\doibase
  10.1016/j.physletb.2016.05.094} {\bibfield  {journal} {\bibinfo  {journal}
  {Phys. Lett.}\ }\textbf {\bibinfo {volume} {B759}},\ \bibinfo {pages}
  {306--312} (\bibinfo {year} {2016}{\natexlab{a}})},\ \Eprint
  {http://arxiv.org/abs/1604.02110} {arXiv:1604.02110 [hep-th]} \BibitemShut
  {NoStop}%
%%CITATION = ARXIV:1604.02110;%%
\bibitem [{\citenamefont {Ma}\ \emph {et~al.}(2017)\citenamefont {Ma},
  \citenamefont {Wang},\ and\ \citenamefont {Yang}}]{Ma:2016vac}%
  \BibitemOpen
  \bibfield  {author} {\bibinfo {author} {\bibfnamefont {Kai}\ \bibnamefont
  {Ma}}, \bibinfo {author} {\bibfnamefont {Jian-Hua}\ \bibnamefont {Wang}}, \
  and\ \bibinfo {author} {\bibfnamefont {Huan-Xiong}\ \bibnamefont {Yang}},\
  }\bibfield  {title} {\enquote {\bibinfo {title} {{Probing the noncommutative
  effects of phase space in the time-dependent Aharonov–Bohm effect}},}\
  }\href {\doibase 10.1016/j.aop.2017.05.005} {\bibfield  {journal} {\bibinfo
  {journal} {Annals Phys.}\ }\textbf {\bibinfo {volume} {383}},\ \bibinfo
  {pages} {120--129} (\bibinfo {year} {2017})},\ \Eprint
  {http://arxiv.org/abs/1605.03902} {arXiv:1605.03902 [hep-ph]} \BibitemShut
  {NoStop}%
%%CITATION = ARXIV:1605.03902;%%
\bibitem [{\citenamefont {Ma}\ \emph {et~al.}(2016{\natexlab{b}})\citenamefont
  {Ma}, \citenamefont {Wang},\ and\ \citenamefont {Yang}}]{Ma:2014tua}%
  \BibitemOpen
  \bibfield  {author} {\bibinfo {author} {\bibfnamefont {Kai}\ \bibnamefont
  {Ma}}, \bibinfo {author} {\bibfnamefont {Jian-hua}\ \bibnamefont {Wang}}, \
  and\ \bibinfo {author} {\bibfnamefont {Huan-Xiong}\ \bibnamefont {Yang}},\
  }\bibfield  {title} {\enquote {\bibinfo {title} {{Seiberg–Witten map and
  quantum phase effects for neutral Dirac particle on noncommutative plane}},}\
  }\href {\doibase 10.1016/j.physletb.2016.03.007} {\bibfield  {journal}
  {\bibinfo  {journal} {Phys. Lett.}\ }\textbf {\bibinfo {volume} {B756}},\
  \bibinfo {pages} {221--227} (\bibinfo {year} {2016}{\natexlab{b}})},\ \Eprint
  {http://arxiv.org/abs/1410.6363} {arXiv:1410.6363 [hep-th]} \BibitemShut
  {NoStop}%
%%CITATION = ARXIV:1410.6363;%%
\bibitem [{\citenamefont {Rodriguez~R.}(2018)}]{Rodriguez:2017iti}%
  \BibitemOpen
  \bibfield  {author} {\bibinfo {author} {\bibfnamefont {Miguel~E.}\
  \bibnamefont {Rodriguez~R.}},\ }\bibfield  {title} {\enquote {\bibinfo
  {title} {{Quantum effects of Aharonov-Bohm type and noncommutative quantum
  mechanics}},}\ }\href {\doibase 10.1103/PhysRevA.97.012109} {\bibfield
  {journal} {\bibinfo  {journal} {Phys. Rev.}\ }\textbf {\bibinfo {volume}
  {A97}},\ \bibinfo {pages} {012109} (\bibinfo {year} {2018})},\ \Eprint
  {http://arxiv.org/abs/1710.10357} {arXiv:1710.10357 [quant-ph]} \BibitemShut
  {NoStop}%
%%CITATION = ARXIV:1710.10357;%%
\bibitem [{\citenamefont {Fateme}\ \emph {et~al.}(2015)\citenamefont {Fateme},
  \citenamefont {Ma},\ and\ \citenamefont {Hassan}}]{Fateme:2015eaa}%
  \BibitemOpen
  \bibfield  {author} {\bibinfo {author} {\bibfnamefont {Hoseini}\ \bibnamefont
  {Fateme}}, \bibinfo {author} {\bibfnamefont {Kai}\ \bibnamefont {Ma}}, \ and\
  \bibinfo {author} {\bibfnamefont {Hassanabadi}\ \bibnamefont {Hassan}},\
  }\bibfield  {title} {\enquote {\bibinfo {title} {{Motion of a Nonrelativistic
  Quantum Particle in Non-commutative Phase Space}},}\ }\href {\doibase
  10.1088/0256-307X/32/10/100302} {\bibfield  {journal} {\bibinfo  {journal}
  {Chin. Phys. Lett.}\ }\textbf {\bibinfo {volume} {32}},\ \bibinfo {pages}
  {100302} (\bibinfo {year} {2015})}\BibitemShut {NoStop}%
%%CITATION = CPLEE,32,100302;%%
\bibitem [{\citenamefont {Wang}\ \emph
  {et~al.}(2018{\natexlab{a}})\citenamefont {Wang}, \citenamefont {Wang},\ and\
  \citenamefont {Ma}}]{Wang:2017gyy}%
  \BibitemOpen
  \bibfield  {author} {\bibinfo {author} {\bibfnamefont {Ya-Hui}\ \bibnamefont
  {Wang}}, \bibinfo {author} {\bibfnamefont {Jian-Hua}\ \bibnamefont {Wang}}, \
  and\ \bibinfo {author} {\bibfnamefont {Kai}\ \bibnamefont {Ma}},\ }\bibfield
  {title} {\enquote {\bibinfo {title} {{Topological Aharonov-Bohm Effect of
  Neutral Scalar Particle on Noncommutative Space}},}\ }\href {\doibase
  10.1007/s10773-017-3627-9} {\bibfield  {journal} {\bibinfo  {journal} {Int.
  J. Theor. Phys.}\ }\textbf {\bibinfo {volume} {57}},\ \bibinfo {pages}
  {951--956} (\bibinfo {year} {2018}{\natexlab{a}})}\BibitemShut {NoStop}%
%%CITATION = IJTPB,57,951;%%
\bibitem [{\citenamefont {Ma}\ and\ \citenamefont {Dulat}(2011)}]{Ma:2011gc}%
  \BibitemOpen
  \bibfield  {author} {\bibinfo {author} {\bibfnamefont {Kai}\ \bibnamefont
  {Ma}}\ and\ \bibinfo {author} {\bibfnamefont {Sayipjamal}\ \bibnamefont
  {Dulat}},\ }\bibfield  {title} {\enquote {\bibinfo {title} {{Spin Hall effect
  on a noncommutative space}},}\ }\href {\doibase 10.1103/PhysRevA.84.012104}
  {\bibfield  {journal} {\bibinfo  {journal} {Phys. Rev.}\ }\textbf {\bibinfo
  {volume} {A84}},\ \bibinfo {pages} {012104} (\bibinfo {year} {2011})},\
  \Eprint {http://arxiv.org/abs/1104.4955} {arXiv:1104.4955 [hep-th]}
  \BibitemShut {NoStop}%
%%CITATION = ARXIV:1104.4955;%%
\bibitem [{\citenamefont {Deriglazov}\ and\ \citenamefont
  {Pupasov-Maksimov}(2016)}]{Deriglazov:2016mhk}%
  \BibitemOpen
  \bibfield  {author} {\bibinfo {author} {\bibfnamefont {Alexei~A.}\
  \bibnamefont {Deriglazov}}\ and\ \bibinfo {author} {\bibfnamefont
  {Andrey~M.}\ \bibnamefont {Pupasov-Maksimov}},\ }\bibfield  {title} {\enquote
  {\bibinfo {title} {{Relativistic corrections to the algebra of position
  variables and spin-orbital interaction}},}\ }\href {\doibase
  10.1016/j.physletb.2016.08.034} {\bibfield  {journal} {\bibinfo  {journal}
  {Phys. Lett.}\ }\textbf {\bibinfo {volume} {B761}},\ \bibinfo {pages}
  {207--212} (\bibinfo {year} {2016})},\ \Eprint
  {http://arxiv.org/abs/1609.00043} {arXiv:1609.00043 [quant-ph]} \BibitemShut
  {NoStop}%
%%CITATION = ARXIV:1609.00043;%%
\bibitem [{\citenamefont {Wang}\ \emph
  {et~al.}(2018{\natexlab{b}})\citenamefont {Wang}, \citenamefont {Long},
  \citenamefont {Long},\ and\ \citenamefont {Wu}}]{Wang:2018zaw}%
  \BibitemOpen
  \bibfield  {author} {\bibinfo {author} {\bibfnamefont {Bing-Qian}\
  \bibnamefont {Wang}}, \bibinfo {author} {\bibfnamefont {Zheng-Wen}\
  \bibnamefont {Long}}, \bibinfo {author} {\bibfnamefont {Chao-Yun}\
  \bibnamefont {Long}}, \ and\ \bibinfo {author} {\bibfnamefont {Shu-Rui}\
  \bibnamefont {Wu}},\ }\bibfield  {title} {\enquote {\bibinfo {title}
  {{Solution of the spin-one DKP oscillator under an external magnetic field in
  noncommutative space with minimal length}},}\ }\href {\doibase
  10.1088/1674-1056/27/1/010301} {\bibfield  {journal} {\bibinfo  {journal}
  {Chin. Phys.}\ }\textbf {\bibinfo {volume} {B27}},\ \bibinfo {pages} {010301}
  (\bibinfo {year} {2018}{\natexlab{b}})}\BibitemShut {NoStop}%
%%CITATION = CHPHD,B27,010301;%%
\bibitem [{\citenamefont {Deriglazov}\ and\ \citenamefont
  {Ramírez}(2016{\natexlab{a}})}]{Deriglazov:2015wde}%
  \BibitemOpen
  \bibfield  {author} {\bibinfo {author} {\bibfnamefont {Alexei~A.}\
  \bibnamefont {Deriglazov}}\ and\ \bibinfo {author} {\bibfnamefont
  {Walberto~Guzmán}\ \bibnamefont {Ramírez}},\ }\bibfield  {title} {\enquote
  {\bibinfo {title} {{Ultrarelativistic Spinning Particle and a Rotating Body
  in External Fields}},}\ }\href {\doibase 10.1155/2016/1376016} {\bibfield
  {journal} {\bibinfo  {journal} {Adv. High Energy Phys.}\ }\textbf {\bibinfo
  {volume} {2016}},\ \bibinfo {pages} {1376016} (\bibinfo {year}
  {2016}{\natexlab{a}})},\ \Eprint {http://arxiv.org/abs/1511.00645}
  {arXiv:1511.00645 [gr-qc]} \BibitemShut {NoStop}%
%%CITATION = ARXIV:1511.00645;%%
\bibitem [{\citenamefont {Deriglazov}\ and\ \citenamefont
  {Guzmán~Ramírez}(2018)}]{Deriglazov:2018vwa}%
  \BibitemOpen
  \bibfield  {author} {\bibinfo {author} {\bibfnamefont {Alexei~A.}\
  \bibnamefont {Deriglazov}}\ and\ \bibinfo {author} {\bibfnamefont {Walberto}\
  \bibnamefont {Guzmán~Ramírez}},\ }\bibfield  {title} {\enquote {\bibinfo
  {title} {{Frame-dragging effect in the field of non rotating body due to unit
  gravimagnetic moment}},}\ }\href {\doibase 10.1016/j.physletb.2018.01.063}
  {\bibfield  {journal} {\bibinfo  {journal} {Phys. Lett.}\ }\textbf {\bibinfo
  {volume} {B779}},\ \bibinfo {pages} {210--213} (\bibinfo {year} {2018})},\
  \Eprint {http://arxiv.org/abs/1802.08079} {arXiv:1802.08079 [gr-qc]}
  \BibitemShut {NoStop}%
%%CITATION = ARXIV:1802.08079;%%
\bibitem [{\citenamefont {Ren}\ and\ \citenamefont {Ma}(2018)}]{Ren:2018rku}%
  \BibitemOpen
  \bibfield  {author} {\bibinfo {author} {\bibfnamefont {Ya-Jie}\ \bibnamefont
  {Ren}}\ and\ \bibinfo {author} {\bibfnamefont {Kai}\ \bibnamefont {Ma}},\
  }\bibfield  {title} {\enquote {\bibinfo {title} {{Influences of the
  coordinate dependent noncommutative space on charged and spin currents}},}\
  }\href {\doibase 10.1142/S0217751X18500938} {\bibfield  {journal} {\bibinfo
  {journal} {Int. J. Mod. Phys.}\ }\textbf {\bibinfo {volume} {A33}},\ \bibinfo
  {pages} {1850093} (\bibinfo {year} {2018})},\ \Eprint
  {http://arxiv.org/abs/1802.10452} {arXiv:1802.10452 [physics.gen-ph]}
  \BibitemShut {NoStop}%
%%CITATION = ARXIV:1802.10452;%%
\bibitem [{\citenamefont {Wang}\ \emph
  {et~al.}(2017{\natexlab{a}})\citenamefont {Wang}, \citenamefont {Zhang},
  \citenamefont {Wang}, \citenamefont {Long},\ and\ \citenamefont
  {Jing}}]{Wang:2017azq}%
  \BibitemOpen
  \bibfield  {author} {\bibinfo {author} {\bibfnamefont {Kang}\ \bibnamefont
  {Wang}}, \bibinfo {author} {\bibfnamefont {Yu-Fei}\ \bibnamefont {Zhang}},
  \bibinfo {author} {\bibfnamefont {Qing}\ \bibnamefont {Wang}}, \bibinfo
  {author} {\bibfnamefont {Zheng-Wen}\ \bibnamefont {Long}}, \ and\ \bibinfo
  {author} {\bibfnamefont {Jian}\ \bibnamefont {Jing}},\ }\bibfield  {title}
  {\enquote {\bibinfo {title} {{Quantum speed limit for a relativistic electron
  in the noncommutative phase space}},}\ }\href@noop {} {\  (\bibinfo {year}
  {2017}{\natexlab{a}})},\ \Eprint {http://arxiv.org/abs/1702.03167}
  {arXiv:1702.03167 [hep-th]} \BibitemShut {NoStop}%
%%CITATION = ARXIV:1702.03167;%%
\bibitem [{\citenamefont {Wang}\ \emph
  {et~al.}(2017{\natexlab{b}})\citenamefont {Wang}, \citenamefont {Zhang},
  \citenamefont {Wang}, \citenamefont {Long},\ and\ \citenamefont
  {Jing}}]{Wang:2017arq}%
  \BibitemOpen
  \bibfield  {author} {\bibinfo {author} {\bibfnamefont {K.}~\bibnamefont
  {Wang}}, \bibinfo {author} {\bibfnamefont {Y.~F.}\ \bibnamefont {Zhang}},
  \bibinfo {author} {\bibfnamefont {Q.}~\bibnamefont {Wang}}, \bibinfo {author}
  {\bibfnamefont {Z.~W.}\ \bibnamefont {Long}}, \ and\ \bibinfo {author}
  {\bibfnamefont {J.}~\bibnamefont {Jing}},\ }\bibfield  {title} {\enquote
  {\bibinfo {title} {{Quantum speed limit for relativistic spin-0 and spin-1
  bosons on commutative and noncommutative planes}},}\ }\href@noop {} {\
  (\bibinfo {year} {2017}{\natexlab{b}})},\ \Eprint
  {http://arxiv.org/abs/1703.01063} {arXiv:1703.01063 [hep-th]} \BibitemShut
  {NoStop}%
%%CITATION = ARXIV:1703.01063;%%
\bibitem [{\citenamefont {Deriglazov}\ and\ \citenamefont
  {Ramírez}(2016{\natexlab{b}})}]{Deriglazov:2015zta}%
  \BibitemOpen
  \bibfield  {author} {\bibinfo {author} {\bibfnamefont {Alexei~A.}\
  \bibnamefont {Deriglazov}}\ and\ \bibinfo {author} {\bibfnamefont
  {Walberto~Guzmán}\ \bibnamefont {Ramírez}},\ }\bibfield  {title} {\enquote
  {\bibinfo {title} {{Mathisson-Papapetrou-Tulczyjew-Dixon (MPTD) equations in
  ultra-relativistic regime and gravimagnetic moment}},}\ }\href {\doibase
  10.1142/S021827181750047X} {\bibfield  {journal} {\bibinfo  {journal} {Int.
  J. Mod. Phys.}\ }\textbf {\bibinfo {volume} {D26}},\ \bibinfo {pages}
  {1750047} (\bibinfo {year} {2016}{\natexlab{b}})},\ \Eprint
  {http://arxiv.org/abs/1509.05357} {arXiv:1509.05357 [gr-qc]} \BibitemShut
  {NoStop}%
%%CITATION = ARXIV:1509.05357;%%
\bibitem [{\citenamefont {Guzmán-Ramírez}\ and\ \citenamefont
  {Deriglazov}(2019)}]{Guzman-Ramirez:2019ijo}%
  \BibitemOpen
  \bibfield  {author} {\bibinfo {author} {\bibfnamefont {Walberto}\
  \bibnamefont {Guzmán-Ramírez}}\ and\ \bibinfo {author} {\bibfnamefont
  {Alexei~A.}\ \bibnamefont {Deriglazov}},\ }\bibfield  {title} {\enquote
  {\bibinfo {title} {{Acceleration of particles in Schwarzschild and Kerr
  geometries}},}\ }\href@noop {} {\  (\bibinfo {year} {2019})},\ \Eprint
  {http://arxiv.org/abs/1902.05450} {arXiv:1902.05450 [gr-qc]} \BibitemShut
  {NoStop}%
%%CITATION = ARXIV:1902.05450;%%
\bibitem [{\citenamefont {Snyder}(1947{\natexlab{a}})}]{Snyder:1946qz}%
  \BibitemOpen
  \bibfield  {author} {\bibinfo {author} {\bibfnamefont {Hartland~S.}\
  \bibnamefont {Snyder}},\ }\bibfield  {title} {\enquote {\bibinfo {title}
  {{Quantized space-time}},}\ }\href {\doibase 10.1103/PhysRev.71.38}
  {\bibfield  {journal} {\bibinfo  {journal} {Phys. Rev.}\ }\textbf {\bibinfo
  {volume} {71}},\ \bibinfo {pages} {38--41} (\bibinfo {year}
  {1947}{\natexlab{a}})}\BibitemShut {NoStop}%
%%CITATION = PHRVA,71,38;%%
\bibitem [{\citenamefont {Snyder}(1947{\natexlab{b}})}]{Snyder:1947nq}%
  \BibitemOpen
  \bibfield  {author} {\bibinfo {author} {\bibfnamefont {Hartland~S.}\
  \bibnamefont {Snyder}},\ }\bibfield  {title} {\enquote {\bibinfo {title}
  {{The Electromagnetic Field in Quantized Space-Time}},}\ }\href {\doibase
  10.1103/PhysRev.72.68} {\bibfield  {journal} {\bibinfo  {journal} {Phys.
  Rev.}\ }\textbf {\bibinfo {volume} {72}},\ \bibinfo {pages} {68--71}
  (\bibinfo {year} {1947}{\natexlab{b}})}\BibitemShut {NoStop}%
%%CITATION = PHRVA,72,68;%%
\bibitem [{\citenamefont {Yang}(1947)}]{Yang:1947ud}%
  \BibitemOpen
  \bibfield  {author} {\bibinfo {author} {\bibfnamefont {C.~N.}\ \bibnamefont
  {Yang}},\ }\bibfield  {title} {\enquote {\bibinfo {title} {{On quantized
  space-time}},}\ }\href {\doibase 10.1103/PhysRev.72.874} {\bibfield
  {journal} {\bibinfo  {journal} {Phys. Rev.}\ }\textbf {\bibinfo {volume}
  {72}},\ \bibinfo {pages} {874} (\bibinfo {year} {1947})}\BibitemShut
  {NoStop}%
%%CITATION = PHRVA,72,874;%%
\bibitem [{\citenamefont {Doplicher}\ \emph {et~al.}(1995)\citenamefont
  {Doplicher}, \citenamefont {Fredenhagen},\ and\ \citenamefont
  {Roberts}}]{Doplicher:1994tu}%
  \BibitemOpen
  \bibfield  {author} {\bibinfo {author} {\bibfnamefont {Sergio}\ \bibnamefont
  {Doplicher}}, \bibinfo {author} {\bibfnamefont {Klaus}\ \bibnamefont
  {Fredenhagen}}, \ and\ \bibinfo {author} {\bibfnamefont {John~E.}\
  \bibnamefont {Roberts}},\ }\bibfield  {title} {\enquote {\bibinfo {title}
  {{The Quantum structure of space-time at the Planck scale and quantum
  fields}},}\ }\href {\doibase 10.1007/BF02104515} {\bibfield  {journal}
  {\bibinfo  {journal} {Commun. Math. Phys.}\ }\textbf {\bibinfo {volume}
  {172}},\ \bibinfo {pages} {187--220} (\bibinfo {year} {1995})},\ \Eprint
  {http://arxiv.org/abs/hep-th/0303037} {arXiv:hep-th/0303037 [hep-th]}
  \BibitemShut {NoStop}%
%%CITATION = HEP-TH/0303037;%%
\bibitem [{\citenamefont {Falomir}\ \emph {et~al.}(2009)\citenamefont
  {Falomir}, \citenamefont {Gamboa}, \citenamefont {Lopez-Sarrion},
  \citenamefont {Mendez},\ and\ \citenamefont {Pisani}}]{Falomir:2009cq}%
  \BibitemOpen
  \bibfield  {author} {\bibinfo {author} {\bibfnamefont {H.}~\bibnamefont
  {Falomir}}, \bibinfo {author} {\bibfnamefont {J.}~\bibnamefont {Gamboa}},
  \bibinfo {author} {\bibfnamefont {J.}~\bibnamefont {Lopez-Sarrion}}, \bibinfo
  {author} {\bibfnamefont {F.}~\bibnamefont {Mendez}}, \ and\ \bibinfo {author}
  {\bibfnamefont {P.~A.~G.}\ \bibnamefont {Pisani}},\ }\bibfield  {title}
  {\enquote {\bibinfo {title} {{Magnetic-Dipole Spin Effects in Noncommutative
  Quantum Mechanics}},}\ }\href {\doibase 10.1016/j.physletb.2009.09.007}
  {\bibfield  {journal} {\bibinfo  {journal} {Phys. Lett.}\ }\textbf {\bibinfo
  {volume} {B680}},\ \bibinfo {pages} {384--386} (\bibinfo {year} {2009})},\
  \Eprint {http://arxiv.org/abs/0905.0157} {arXiv:0905.0157 [hep-th]}
  \BibitemShut {NoStop}%
%%CITATION = ARXIV:0905.0157;%%
\bibitem [{\citenamefont {Das}\ \emph {et~al.}(2011)\citenamefont {Das},
  \citenamefont {Falomir}, \citenamefont {Nieto}, \citenamefont {Gamboa},\ and\
  \citenamefont {Mendez}}]{Das:2011tj}%
  \BibitemOpen
  \bibfield  {author} {\bibinfo {author} {\bibfnamefont {Ashok}\ \bibnamefont
  {Das}}, \bibinfo {author} {\bibfnamefont {H.}~\bibnamefont {Falomir}},
  \bibinfo {author} {\bibfnamefont {M.}~\bibnamefont {Nieto}}, \bibinfo
  {author} {\bibfnamefont {J.}~\bibnamefont {Gamboa}}, \ and\ \bibinfo {author}
  {\bibfnamefont {F.}~\bibnamefont {Mendez}},\ }\bibfield  {title} {\enquote
  {\bibinfo {title} {{Aharonov-Bohm effect in a Class of Noncommutative
  Theories}},}\ }\href {\doibase 10.1103/PhysRevD.84.045002} {\bibfield
  {journal} {\bibinfo  {journal} {Phys. Rev.}\ }\textbf {\bibinfo {volume}
  {D84}},\ \bibinfo {pages} {045002} (\bibinfo {year} {2011})},\ \Eprint
  {http://arxiv.org/abs/1105.1800} {arXiv:1105.1800 [hep-th]} \BibitemShut
  {NoStop}%
%%CITATION = ARXIV:1105.1800;%%
\bibitem [{\citenamefont {Guzmán~Ramírez}\ \emph {et~al.}(2014)\citenamefont
  {Guzmán~Ramírez}, \citenamefont {Deriglazov},\ and\ \citenamefont
  {Pupasov-Maksimov}}]{Ramirez:2013xga}%
  \BibitemOpen
  \bibfield  {author} {\bibinfo {author} {\bibfnamefont {Walberto}\
  \bibnamefont {Guzmán~Ramírez}}, \bibinfo {author} {\bibfnamefont {Alexei~A.}\
  \bibnamefont {Deriglazov}}, \ and\ \bibinfo {author} {\bibfnamefont
  {Andrey~M.}\ \bibnamefont {Pupasov-Maksimov}},\ }\bibfield  {title} {\enquote
  {\bibinfo {title} {{Frenkel electron and a spinning body in a curved
  background}},}\ }\href {\doibase 10.1007/JHEP03(2014)109} {\bibfield
  {journal} {\bibinfo  {journal} {JHEP}\ }\textbf {\bibinfo {volume} {03}},\
  \bibinfo {pages} {109} (\bibinfo {year} {2014})},\ \Eprint
  {http://arxiv.org/abs/1311.5743} {arXiv:1311.5743 [hep-th]} \BibitemShut
  {NoStop}%
%%CITATION = ARXIV:1311.5743;%%
\bibitem [{\citenamefont {Deriglazov}\ and\ \citenamefont
  {Ramírez}(2015)}]{Deriglazov:2015bqa}%
  \BibitemOpen
  \bibfield  {author} {\bibinfo {author} {\bibfnamefont {Alexei~A.}\
  \bibnamefont {Deriglazov}}\ and\ \bibinfo {author} {\bibfnamefont
  {Walberto~Guzmán}\ \bibnamefont {Ramírez}},\ }\bibfield  {title} {\enquote
  {\bibinfo {title} {{Lagrangian formulation for
  Mathisson-Papapetrou-Tulczyjew-Dixon (MPTD) equations}},}\ }\href {\doibase
  10.1103/PhysRevD.92.124017} {\bibfield  {journal} {\bibinfo  {journal} {Phys.
  Rev.}\ }\textbf {\bibinfo {volume} {D92}},\ \bibinfo {pages} {124017}
  (\bibinfo {year} {2015})},\ \Eprint {http://arxiv.org/abs/1509.04926}
  {arXiv:1509.04926 [gr-qc]} \BibitemShut {NoStop}%
%%CITATION = ARXIV:1509.04926;%%
\bibitem [{\citenamefont {Ferrari}\ \emph {et~al.}(2013)\citenamefont
  {Ferrari}, \citenamefont {Gomes}, \citenamefont {Kupriyanov},\ and\
  \citenamefont {Stechhahn}}]{Ferrari:2012bv}%
  \BibitemOpen
  \bibfield  {author} {\bibinfo {author} {\bibfnamefont {A.~F.}\ \bibnamefont
  {Ferrari}}, \bibinfo {author} {\bibfnamefont {M.}~\bibnamefont {Gomes}},
  \bibinfo {author} {\bibfnamefont {V.~G.}\ \bibnamefont {Kupriyanov}}, \ and\
  \bibinfo {author} {\bibfnamefont {C.~A.}\ \bibnamefont {Stechhahn}},\
  }\bibfield  {title} {\enquote {\bibinfo {title} {{Dynamics of a Dirac Fermion
  in the presence of spin noncommutativity}},}\ }\href {\doibase
  10.1016/j.physletb.2012.12.010} {\bibfield  {journal} {\bibinfo  {journal}
  {Phys. Lett.}\ }\textbf {\bibinfo {volume} {B718}},\ \bibinfo {pages}
  {1475--1480} (\bibinfo {year} {2013})},\ \Eprint
  {http://arxiv.org/abs/1207.0412} {arXiv:1207.0412 [hep-th]} \BibitemShut
  {NoStop}%
%%CITATION = ARXIV:1207.0412;%%
\bibitem [{\citenamefont {Huang}\ and\ \citenamefont
  {Wang}(2013)}]{Huang:2012kf}%
  \BibitemOpen
  \bibfield  {author} {\bibinfo {author} {\bibfnamefont {Jia-Hui}\ \bibnamefont
  {Huang}}\ and\ \bibinfo {author} {\bibfnamefont {Weijian}\ \bibnamefont
  {Wang}},\ }\bibfield  {title} {\enquote {\bibinfo {title} {{Microcausality of
  spin-induced noncommutative theories}},}\ }\href {\doibase
  10.1142/S0217732312502392} {\bibfield  {journal} {\bibinfo  {journal} {Mod.
  Phys. Lett.}\ }\textbf {\bibinfo {volume} {A28}},\ \bibinfo {pages} {1250239}
  (\bibinfo {year} {2013})},\ \Eprint {http://arxiv.org/abs/1203.1992}
  {arXiv:1203.1992 [hep-th]} \BibitemShut {NoStop}%
%%CITATION = ARXIV:1203.1992;%%
\bibitem [{\citenamefont {Vasyuta}\ and\ \citenamefont
  {Tkachuk}(2016)}]{Vasyuta:2016gyz}%
  \BibitemOpen
  \bibfield  {author} {\bibinfo {author} {\bibfnamefont {V.~M.}\ \bibnamefont
  {Vasyuta}}\ and\ \bibinfo {author} {\bibfnamefont {V.~M.}\ \bibnamefont
  {Tkachuk}},\ }\bibfield  {title} {\enquote {\bibinfo {title} {{Classical
  electrodynamics in a space with spin noncommutativity of coordinates}},}\
  }\href {\doibase 10.1016/j.physletb.2016.09.001} {\bibfield  {journal}
  {\bibinfo  {journal} {Phys. Lett.}\ }\textbf {\bibinfo {volume} {B761}},\
  \bibinfo {pages} {462--468} (\bibinfo {year} {2016})},\ \Eprint
  {http://arxiv.org/abs/1604.07448} {arXiv:1604.07448 [hep-th]} \BibitemShut
  {NoStop}%
%%CITATION = ARXIV:1604.07448;%%
\bibitem [{\citenamefont {Greiner}(1990)}]{Greiner:1990tz}%
  \BibitemOpen
  \bibfield  {author} {\bibinfo {author} {\bibfnamefont {W.}~\bibnamefont
  {Greiner}},\ }\href@noop {} {\emph {\bibinfo {title} {{Relativistic quantum
  mechanics: Wave equations}}}}\ (\bibinfo {year} {1990})\BibitemShut {NoStop}%
%%CITATION = INSPIRE-307769;%%
\bibitem [{\citenamefont {Bagchi}\ and\ \citenamefont
  {Fring}(2009)}]{Bagchi:2009wb}%
  \BibitemOpen
  \bibfield  {author} {\bibinfo {author} {\bibfnamefont {Bijan}\ \bibnamefont
  {Bagchi}}\ and\ \bibinfo {author} {\bibfnamefont {Andreas}\ \bibnamefont
  {Fring}},\ }\bibfield  {title} {\enquote {\bibinfo {title} {{Minimal length
  in Quantum Mechanics and non-Hermitian Hamiltonian systems}},}\ }\href
  {\doibase 10.1016/j.physleta.2009.09.054} {\bibfield  {journal} {\bibinfo
  {journal} {Phys. Lett.}\ }\textbf {\bibinfo {volume} {A373}},\ \bibinfo
  {pages} {4307--4310} (\bibinfo {year} {2009})},\ \Eprint
  {http://arxiv.org/abs/0907.5354} {arXiv:0907.5354 [hep-th]} \BibitemShut
  {NoStop}%
%%CITATION = ARXIV:0907.5354;%%
\bibitem [{\citenamefont {Fring}\ \emph {et~al.}(2010)\citenamefont {Fring},
  \citenamefont {Gouba},\ and\ \citenamefont {Scholtz}}]{Fring:2010pw}%
  \BibitemOpen
  \bibfield  {author} {\bibinfo {author} {\bibfnamefont {Andreas}\ \bibnamefont
  {Fring}}, \bibinfo {author} {\bibfnamefont {Laure}\ \bibnamefont {Gouba}}, \
  and\ \bibinfo {author} {\bibfnamefont {Frederik~G.}\ \bibnamefont
  {Scholtz}},\ }\bibfield  {title} {\enquote {\bibinfo {title} {{Strings from
  position-dependent noncommutativity}},}\ }\href {\doibase
  10.1088/1751-8113/43/34/345401} {\bibfield  {journal} {\bibinfo  {journal}
  {J. Phys.}\ }\textbf {\bibinfo {volume} {A43}},\ \bibinfo {pages} {345401}
  (\bibinfo {year} {2010})},\ \Eprint {http://arxiv.org/abs/1003.3025}
  {arXiv:1003.3025 [hep-th]} \BibitemShut {NoStop}%
%%CITATION = ARXIV:1003.3025;%%
\bibitem [{\citenamefont {Tanabashi}\ \emph {et~al.}(2018)\citenamefont
  {Tanabashi} \emph {et~al.}}]{Tanabashi:2018oca}%
  \BibitemOpen
  \bibfield  {author} {\bibinfo {author} {\bibfnamefont {M.}~\bibnamefont
  {Tanabashi}} \emph {et~al.} (\bibinfo {collaboration} {Particle Data
  Group}),\ }\bibfield  {title} {\enquote {\bibinfo {title} {{Review of
  Particle Physics}},}\ }\href {\doibase 10.1103/PhysRevD.98.030001} {\bibfield
   {journal} {\bibinfo  {journal} {Phys. Rev.}\ }\textbf {\bibinfo {volume}
  {D98}},\ \bibinfo {pages} {030001} (\bibinfo {year} {2018})}\BibitemShut
  {NoStop}%
%%CITATION = PHRVA,D98,030001;%%
\bibitem [{\citenamefont {Hagiwara}\ \emph {et~al.}(2018)\citenamefont
  {Hagiwara}, \citenamefont {Ma},\ and\ \citenamefont
  {Mukhopadhyay}}]{Hagiwara:2017lse}%
  \BibitemOpen
  \bibfield  {author} {\bibinfo {author} {\bibfnamefont {Kaoru}\ \bibnamefont
  {Hagiwara}}, \bibinfo {author} {\bibfnamefont {Kai}\ \bibnamefont {Ma}}, \
  and\ \bibinfo {author} {\bibfnamefont {Satyanarayan}\ \bibnamefont
  {Mukhopadhyay}},\ }\bibfield  {title} {\enquote {\bibinfo {title} {{Closing
  in on the chargino contribution to the muon g-2 in the MSSM: current LHC
  constraints}},}\ }\href {\doibase 10.1103/PhysRevD.97.055035} {\bibfield
  {journal} {\bibinfo  {journal} {Phys. Rev.}\ }\textbf {\bibinfo {volume}
  {D97}},\ \bibinfo {pages} {055035} (\bibinfo {year} {2018})},\ \Eprint
  {http://arxiv.org/abs/1706.09313} {arXiv:1706.09313 [hep-ph]} \BibitemShut
  {NoStop}%
%%CITATION = ARXIV:1706.09313;%%
\end{thebibliography}%

\end{document}